\documentclass[prd,twocolumn,showpacs,nofootinbib,amsmath,amssymb]{revtex4}

\usepackage{graphicx,bm}

\newcommand{\Mmax}{M_{\mathrm{max}}}
\newcommand{\Mmin}{M_{\mathrm{min}}}
\newcommand{\Lmin}{L_{\mathrm{min}}}
\newcommand{\Lmax}{L_{\mathrm{max}}}

\def\VEV#1{\left\langle #1 \right\rangle}

\begin{document}

\title{Gamma-ray background anisotropy from galactic dark matter
substructure}

\author{Shin'ichiro Ando}
%\email{ando@tapir.caltech.edu}
\affiliation{California Institute of Technology, Mail Code 350-17,
Pasadena, California 91125, USA}

\date{March 26, 2009; accepted July 1, 2009}

\begin{abstract}
Dark matter annihilation in galactic substructure would imprint
 characteristic angular signatures on the all-sky map of the diffuse
 gamma-ray background.
We study the gamma-ray background anisotropy due to the subhalos and
 discuss detectability at Fermi Gamma-ray Space Telescope.
In contrast to earlier work that relies on simulated all-sky maps, we
 derive analytic formulae that enable to directly compute the angular
 power spectrum, given parameters of subhalos such as mass function and
 radial profile of gamma-ray luminosity.
As our fiducial subhalo models, we adopt $M^{-1.9}$ mass spectrum,
 subhalos radial distribution suppressed towards the galactic center,
 and luminosity profile of each subhalo dominated by its smooth
 component.
We find that, for multipole regime corresponding to $\theta
 \alt 5^\circ$, the angular power spectrum is dominated by a noise-like
 term, with suppression due to internal structure of relevant subhalos.
If the mass spectrum extends down to Earth-mass scale, then the
 subhalos would be detected in the anisotropy with Fermi at angular
 scales of $\sim$10$^\circ$, if their contribution to the gamma-ray
 background is larger than $\sim$20\%.
If the minimum mass is around $10^4 M_\odot$, on the other hand, the
 relevant angular scale for detection is $\sim$1$^\circ$, and the
 anisotropy detection requires that the subhalo contribution to
 the gamma-ray background intensity is only $\sim$4\%.
These can be achieved with a modest boost for particle physics
 parameters.
We also find that the anisotropy analysis could be a more sensitive
 probe for the subhalos than individual detection.
We also study dependence on model parameters, where we reach the similar
 conclusions for all the models investigated.
The analytic approach should be very useful when Fermi data are analyzed
 and the obtained angular power spectrum is interpreted in terms of
 subhalo models.
\end{abstract}


\maketitle

\section{Introduction}
\label{sec:introduction}

Modern astrophysical and cosmological measurements strongly support
existence of nonbaryonic dark matter.
Although the true identity of dark matter is unknown observationally and
experimentally, there are several well-motivated particle-physics models
that provide a candidate particle for dark matter.
Weakly interacting massive particles (WIMPs) such as supersymmetric
neutralinos are perhaps the most popular candidate~\cite{Jungman1996,
Hooper2007a}.
As they interact with themselves as well as standard-model particles,
many experiments are being carried out to look for signatures of
scattering of WIMPs off nuclei in underground detectors and of
self-annihilation of WIMP particles in dark matter
halos~\cite{Bertone2005}.

Recently launched Fermi Gamma-ray Space Telescope~\cite{Atwood2009}
features promising capability to detect gamma rays from WIMP
annihilation in the right energy range~\cite{Baltz2008}.
In addition, recent numerical simulations find that dark matter in a
host halo is distributed in clumpy substructure
(subhalos)~\cite{Moore1999, Klypin1999, Diemand2005, Diemand2006}
whose masses range quite widely, potentially down to Earth
mass~\cite{Hofmann2001, Green2005, Loeb2005, Profumo2006,
Bertschinger2006}.
This feature is encouraging because annihilation probability is
proportional to density squared, and thus the significant clumpiness
boosts the gamma-ray yields.
Several avenues have been proposed for Fermi to look for annihilation
gamma rays: galactic center~\cite{Berezinsky1994, Bergstrom1998,
Cesarini2004, Fornengo2004, Dodelson2008}, relatively massive
substructure often associated with dwarf galaxies~\cite{Bergstrom1999,
Calcaneo-Roldan2000, Tasitsiomi2002, Stoehr2003, Evans2004, Aloisio2004,
Koushiappas2004, Diemand2007a, Strigari2008, Pieri2008, Kuhlen2008,
Martinez2009}, proper motions of nearby small
subhalos~\cite{Koushiappas2006, Ando2008}, diffuse gamma-ray
background~\cite{Bergstrom2001, Ullio2002, Taylor2003, Elsaesser2005,
Ando2005, Oda2005, Horiuchi2006}, etc.

There is an increasing interest in statistical analysis of the all-sky
map of the gamma-ray background obtained with Fermi in the near future.
References~\cite{Ando2006, Ando2007a} computed angular power spectrum of
the gamma-ray background from annihilation in extragalactic dark matter
halos, and showed that the signature would be different from that of
ordinary astrophysical sources (see also, Refs.~\cite{Cuoco2007,
Cuoco2008, Zhang2008, Taoso2008}).
The same approach has been applied to signals from substructure in
the galactic halo~\cite{Siegal-Gaskins2008, Fornasa2009}.
This is indeed important, because the galactic substructure typically
gives a larger contribution to the diffuse gamma-ray background than
extragalactic halos do~\cite{Fornasa2009, Hooper2007b, Yuksel2007}.
In Refs.~\cite{Siegal-Gaskins2008, Fornasa2009}, first the all-sky
gamma-ray map was simulated from sets of subhalo models, and then the
map was analyzed to obtain the angular power spectrum.
In addition, one-point probability distribution function of the
gamma-ray flux has also been studied~\cite{Lee2008, Dodelson2009}.

In this paper, we revisit the gamma-ray background anisotropy from
dark matter annihilation in the galactic subhalos.
In contrast to the earlier works~\cite{Siegal-Gaskins2008, Fornasa2009}
that heavily relied on mock gamma-ray maps generated from subhalo
models, we develop an analytic approach to compute the angular power
spectrum directly.
This way, we are able to calculate the angular power spectrum easily and
more quickly, if we specify some input parameters and characteristics of
galactic subhalos.
This would be in particular useful when we have results of actual Fermi
data analysis, and try to give physical interpretation for them.

We find that the angular power spectrum $C_\ell$ is divided into two
parts: one depending on (ensemble-averaged) distribution of
subhalos in a host (``two-subhalo'' term, $C_\ell^{\rm 2sh}$), and the
other depending on the emissivity profile of single subhalos as well as
the number of subhalos significantly contributing to the background
intensity (``one-subhalo'' term, $C_\ell^{\rm 1sh}$).
The latter would be shot noise if the subhalos were completely
point sources, but Fermi will be able to see deviations from the shot
noise due to the angular extension of the relevant subhalos.
Using the latest subhalo models following recent numerical
simulations, we give predictions for the angular power spectrum, and
show that the multipole range $10 \alt \ell \alt 100$ would be a
favorable window for anisotropy detection.
We also discuss the detectability of the angular power spectrum from
subhalos with Fermi, which turns out to be promising, potentially better
than the detection of subhalos as identified gamma-ray sources.

This paper is organized as follows.
In Sec.~\ref{sec:formulation}, we give formulation for angular power
spectrum as well as mean intensity of the gamma-ray background.
These formulae derived are applied to several subhalo models in the
subsequent sections.
In Sec.~\ref{sec:point source}, we study a simple case in which all the
subhalos are assumed to be a point-like gamma-ray emitters.
The case of extended subhalos is addressed in Sec.~\ref{sec:realistic
case}, where we also discuss detectability with Fermi.
We close this paper by discussing the results in
Sec.~\ref{sec:discussion} and by giving concluding remarks in
Sec.~\ref{sec:conclusions}.

\section{Formulation}
\label{sec:formulation}

\subsection{Relevant quantities of subhalos}

We assume that a fraction $f$ of the mass of the galactic halo is in the
form of subhalos, and $1-f$ is distributed as a smooth halo.
For the density profile of the Milky-Way dark matter halo, we adopt a
spherically symmetric Navarro-Frenk-White (NFW)
profile~\cite{Navarro1997}:
\begin{equation}
 \rho_{\rm MW, NFW}(r) = \frac{\rho_{s,{\rm MW}}}{(r/r_{s,{\rm MW}})(1 +
 r/r_{s, {\rm MW}})^2},
\end{equation}
where $r$ is the galactocentric radius, $r_{s,{\rm MW}}$ and
$\rho_{s,{\rm MW}}$ are the scale radius and scale density of the
Milky-Way halo, respectively.
This profile extends up to a virial radius $r_{\rm vir, MW}$, and an
enclosed mass within this radius is defined as a virial mass $M_{\rm
vir, MW}$.
We use the following values for these parameters: $r_{s, {\rm MW}} =
21.5$ kpc, $\rho_{s, {\rm MW}} = 4.9 \times 10^6 M_{\odot}$ kpc$^{-3}$,
$r_{\rm vir, MW} = 258$ kpc, and $M_{\rm vir, MW} = 10^{12}
M_{\odot}$~\cite{Klypin2002}.
% (but our results are largely
%independent of the choice of these parameters).

%Radial distribution of subhalos follows the density profile of the
%host to some extent.
%Furthermore, subhalos masses are likely to range quite widely as implied
%by numerical simulations.
As subhalo number density per unit mass range, we define a
subhalo mass function, $dn_{\rm sh} (r, M) / dM$, and upper and lower
limits of the function by $\Mmax$ and $\Mmin$, respectively.
Numerical simulations imply that the shape of mass distribution follows
typically a power law, which is close to $dn_{\rm sh} / dM \propto
M^{-2}$, i.e., the same amount of subhalo masses per
decade~\cite{Ghigna2000, DeLucia2004, Gao2004, Shaw2007,
Diemand2007b}.
We also assume that there is a one-to-one relation between subhalo
masses and luminosities $L(M)$, and therefore, the luminosity function
is written as $dn_{\rm sh} / dL = (dn_{\rm sh}/dM) |dM/dL|$.
The upper and lower limits on the luminosity function are then given by
$\Lmax = L(\Mmax)$ and $\Lmin = L(\Mmin)$, respectively.
After integrating the mass (luminosity) function over mass (luminosity),
we obtain the number density of subhalos $n_{\rm sh}(r)$.

We assume that each subhalo has extended, isotropic emissivity profile
around its center (we call this ``seed'' position), $u(r_{\rm sh},M) L$,
with the profile function $u(r_{\rm sh}, M)$ normalized so that it
gives unity after volume integration.
We also define the Fourier transform of $u(r_{\rm sh}, M)$: $\tilde u(k,
M)$, where $k$ is the wave number.

The most relevant equations are Eqs.~(\ref{eq:intensity}) and
(\ref{eq:angular power spectrum})--(\ref{eq:angular power spectrum
2sh}) derived in the remainder of this section.
The readers who are only interested in application of these equations to
subhalo models may skip to Sec.~\ref{sec:point source}.

\subsection{Gamma-ray intensity from subhalos}
\label{sub:intensity}

We label positions of seed of a subhalo $i$ by $\bm x_i$, and its
luminosity and mass by $L_i$ and $M_i$, respectively.
With these definitions, the gamma-ray intensity towards a direction
$\hat{\bm n}$ is given by the line-of-sight integration ($ds$) of the
emissivity:
\begin{eqnarray}
I(\hat{\bm n}) &=& \frac{1}{4\pi} \int ds \sum_i u(s\hat{\bm n} - \bm
 x_i, M_i) L_i
 \nonumber\\&=&
 \frac{1}{4\pi} \int ds \int dL \int d^3 x \sum_i
 \delta^3 (\bm x - \bm x_i) \delta (L - L_i)
 \nonumber\\&&{}\times
 u(s \hat{\bm n} - \bm x, M) L,
\label{eq:bare intensity}
\end{eqnarray}
for one realization of the universe, where $\delta^N$ is the
$N$-dimensional delta function.
Throughout this paper, we define the intensity as a number of gamma-ray
photons per unit area, time, and solid angle, and the luminosity as a
number of photons emitted per unit time.
We also assume $E \ge 10$ GeV as a targeted gamma-ray energy.

We now take ensemble average over infinite number of realizations of the
universe.
The discrete source distribution then becomes continuous function; i.e.,
\begin{equation}
\VEV{\sum_i \delta^3(\bm x - \bm x_i) \delta(L - L_i)} =
 \frac{dn_{\rm sh}(\bm x, L)}{dL},
\end{equation}
where the bracket represents the ensemble average.
Using these in Eq.~(\ref{eq:bare intensity}), an ensemble-averaged
intensity is
\begin{eqnarray}
\VEV{I(\hat{\bm n})} &=&
 \frac{1}{4\pi} \int ds \int dL \int d^3 x
\frac{dn_{\rm sh}(\bm x, L)}{dL}
 \nonumber\\&&{}\times
 u(s \hat{\bm n} - \bm x, M) L.
\end{eqnarray}
We further assume that spatial extension of each subhalo is much smaller
than a scale on which subhalo distribution significantly changes.
With this reasonable assumption, we could take the luminosity function
$dn_{\rm sh}/dL$ out of the volume integral by taking $\bm x \approx
s\hat{\bm n}$, since it is a slowly varying function of $\bm x$.
As the integration of $u(s \hat{\bm n} - \bm x, M)$ over $\bm x$
simply becomes unity, the ensemble average of intensity is
\begin{equation}
\VEV{I(\hat{\bm n})} = \frac{1}{4\pi} \int_{\Lmin}^{\Lmax}dL
 \int_{s_\ast(L)}^{s_{\rm max}(\hat{\bm n})} ds \frac{dn_{\rm
 sh}(r[s,\hat{\bm n}],L)}{dL} L,
\label{eq:intensity}
\end{equation}
where we specified upper and lower limits of the integrals.
Galactocentric radius corresponding to $s\hat{\bm n}$ (appearing as
index of luminosity function) is obtained through the relation: $r^2 =
r_\odot^2 + s^2 - 2 r_\odot s \cos\psi$, where $r_\odot = 8.5$ kpc is
the galactocentric radius of the solar system and $\psi$ is the angle
between $\hat{\bm n}$ and the direction to the galactic center.
We obtain the lower limit of the $s$-integral by the detection criterion
$L = 4\pi s_\ast^2 F_{\rm sens}$ with the flux sensitivity of Fermi
(typically $F_{\rm sens} \simeq 2\times 10^{-10}$ cm$^{-2}$ s$^{-1}$ for
photons that we consider).
By setting this, we do not add contributions from subhalos bright enough
to be identified as individual sources.
The upper limit $s_{\rm max} (\hat{\bm n})$ corresponds to $r_{\rm vir,
MW}$ through the relation $r_{\rm vir, MW}^2 = r_{\odot}^2 + s_{\rm
max}^2 - 2 r_{\odot} s_{\rm max} \cos \psi$.

By further averaging over directions $\hat{\bm n}$, we obtain a mean
gamma-ray intensity
\begin{eqnarray}
\overline{\VEV{I}} &=& \frac{1}{\Omega_{\rm sky}} \int
 d\Omega_{\hat{\bm n}} \VEV{I(\hat{\bm n})}
\nonumber\\&\approx&
 \frac{1}{4\pi} \int_{\Lmin}^{\Lmax}dL
 \int_{s_\ast(L)}^{r_{\rm vir,MW}} ds \frac{d\overline{n_{\rm
 sh}}(s,L)}{dL} L,
\label{eq:mean intensity}
\end{eqnarray}
where we represent quantities averaged over $\hat{\bm n}$ by putting a
horizontal line on top of them, and $\Omega_{\rm sky}$ is the solid
angle of the sky over which the averages are taken.
In the second equality, we approximate $s_{\rm max} \approx r_{\rm vir,
MW}$, as $r_{\odot} \ll r_{\rm vir, MW}$; since $s_{\rm max}$ is now
independent of $\hat{\bm n}$, we could perform the angle average of
$dn_{\rm sh}/dL$ before integrating over $s$ and $L$.

The expected number of subhalos that would be detected as individual
sources from $\Omega_{\rm sky}$ is then
\begin{equation}
\VEV{N_{\rm sh}} = \Omega_{\rm sky}
 \int_{\Lmin}^{\Lmax}dL \int_0^{s_\ast (L)} ds s^2
 \frac{d\overline{n_{\rm sh}} (s,L)}{dL},
\label{eq:number of detectable subhalos}
\end{equation}
as we count sources close enough to give a flux larger than $F_{\rm
sens}$; i.e., $s < s_\ast(L)$.

\subsection{Angular power spectrum of gamma-ray background from
  subhalos}
\label{sub:angular power spectrum}

We decompose the gamma-ray intensity map with spherical harmonics:
\begin{equation}
\frac{\delta I(\hat{\bm n})}{\overline{\VEV{I}}} \equiv
 \frac{I(\hat{\bm n}) - \overline{\VEV{I}}}{\overline{\VEV{I}}} =
 \sum_{\ell m} a_{\ell m}Y_{\ell m} (\hat{\bm n}).
\end{equation}
Therefore (dimensionless) expansion coefficients $a_{\ell m}$ are
obtained from the intensity map through the inverse relation,
\begin{eqnarray}
a_{\ell m} & \simeq & \frac{1}{\overline{\VEV{I}}f_{\rm sky}}
 \int d\Omega_{\hat{\bm n}} \delta
 I(\hat{\bm n}) Y_{\ell m}^\ast (\hat{\bm n})
 \nonumber\\& = &
 \frac{1}{\overline{\VEV{I}}f_{\rm sky}} \int d\Omega_{\hat{\bm n}}
 I(\hat{\bm n}) Y_{\ell m}^\ast (\hat{\bm n}),
\label{eq:a_lm}
\end{eqnarray}
where $f_{\rm sky} \equiv \Omega_{\rm sky} / 4\pi$, and the second
equality holds except for monopole (i.e., for $\ell \neq 0$).

From Eq.~(\ref{eq:a_lm}), we have, for $\ell \ge 1$,
\begin{eqnarray}
\VEV{|a_{\ell m}|^2} & = &
 \frac{1}{f_{\rm sky}^2}\int d\Omega_{\hat{\bm n}_1} \int
 d\Omega_{\hat{\bm n}_2} C(\hat{\bm n}_1, \hat{\bm n}_2)
 \nonumber\\&&{}\times
 Y_{\ell m}^\ast(\hat{\bm n}_1) Y_{\ell m}(\hat{\bm n}_2).
\label{eq:a_lm squared}
\end{eqnarray}
where $C(\hat{\bm n}_1, \hat{\bm n}_2) \equiv \VEV{I(\hat{\bm
 n}_1) I(\hat{\bm n}_2)} / \overline{\VEV{I}}^2$.
Thus, we want to evaluate $\VEV{I(\hat{\bm n}_1) I(\hat{\bm n}_2)}$,
which now through Eq.~(\ref{eq:bare intensity}) depends on
\begin{eqnarray}
\lefteqn{\VEV{\sum_{i,j} \delta^3(\bm x_1 - \bm x_i) \delta (L_1 - L_i)
 \delta^3(\bm x_2 - \bm x_j) \delta (L_2 - L_j)}}
\hspace{1.2cm}\nonumber\\
&=&
 \frac{dn_{\rm sh}(\bm x_1, L_1)}{dL}
 \frac{dn_{\rm sh}(\bm x_2, L_2)}{dL}
 (1+\xi_{\rm sh})
 \nonumber\\&&{}
 +\frac{dn_{\rm sh}(\bm x_1, L_1)}{dL} \delta^3(\bm x_1 - \bm x_2)
 \delta(L_1 - L_2).
\nonumber\\
\label{eq:one and two-subhalo terms}
\end{eqnarray}
Since $\bm x_1$ and $\bm x_2$ represent positions of the subhalo seeds,
the first term correlates positions and luminosities of two distinct
subhalos (two-subhalo term).
Here $\xi_{\rm sh}$ is the intrinsic two-point correlation function of
the subhalo seeds.
% averaged over the volume of the galactic halo
This two-subhalo term corresponds to a ``one-halo term'' of the halo
model~\cite{Seljak2000}, which is proportional to $n_{\rm sh}(\bm x_1)
n_{\rm sh}(\bm x_2) \VEV{N(N-1)}$ and gives dominant contribution to the
galaxy power spectrum at scales smaller than the virial radius of
halos.
Although in the halo model one often discusses {\it galaxy} power
spectrum and $N$ stands for the total number of galaxies in the host
halo, exactly the same argument can be applied to the {\it subhalo}
power spectrum, and therefore, $N$ is regarded as number of subhalos
instead.
The correlation function $\xi_{\rm sh}$ is related to $N$ via
$1+\xi_{\rm sh} = \VEV{N(N-1)} / \VEV{N}^2$, and numerical simulations
show that $1 + \xi_{\rm sh}$ is very close to 1, if the host is massive
enough to contain large number of galaxies (subhalos), $N \gg
1$~\cite{Kravtsov2004, Zentner2005}, the case we consider here.
The second term of Eq.~(\ref{eq:one and two-subhalo terms}), on the
other hand, represents the case of one identical halo where $\bm x_1 =
\bm x_2$ and $L_1 = L_2$ (one-subhalo term).
Therefore, one-subhalo and two-subhalo terms of $\VEV{I(\hat{\bm n_1})
I(\hat{\bm n}_2)}$ are
\begin{eqnarray}
C_{\rm 1sh}(\hat{\bm n}_1,\hat{\bm n}_2)&=&
 \frac{1}{16 \pi^2 \overline{\VEV{I}}^2}\int ds_1 \int ds_2
 \int dL \int d^3 x L^2
 \nonumber\\&&{}\times
 \frac{dn_{\rm sh}(\bm x, L)}{dL}
 u(s_1\hat{\bm n}_1-\bm x,M)
 \nonumber\\&&{}\times
 u(s_2\hat{\bm n}_2-\bm x,M),
 \label{eq:correlation 1sh} \\
C_{\rm 2sh}(\hat{\bm n}_1, \hat{\bm n}_2) &=&
 \frac{\VEV{I(\hat{\bm n}_1)} \VEV{I(\hat{\bm n}_2)}}
 {\overline{\VEV{I}}^2} (1+\xi_{\rm sh}).
 \label{eq:correlation 2sh}
\end{eqnarray}

We work on to further simplify Eq.~(\ref{eq:correlation 1sh}).
First, we again take the luminosity function out of $\bm x$-integral
with the index $\bm x \approx s_1 \hat{\bm n}_1 \approx s_2 \hat{\bm
n}_2$, as its change on subhalo scales is moderate.
Then, we take Fourier transforms of $u(r_{\rm sh},M)$: $\tilde u(k,M)$.
With this, $\bm x$-integral gives the delta function that collapses one
of the $\bm k$-integrals, leaving a common wave number.
Finally, we use the small-separation approximation (e.g.,
\cite{Peebles1980}), where we take $s$, $\eta$, $\theta$, $\hat{\bm n}$,
and $\hat{\bm \theta}$ instead of $s_1$, $s_2$, $\hat{\bm n}_1$, and
$\hat{\bm n}_2$; these quantities are related via $s = (s_1 + s_2) / 2$,
$\eta = s_2 - s_1$, $\cos\theta = \hat{\bm n}_1 \cdot \hat{\bm n}_2$ and
$s_1 \hat{\bm n}_1 - s_2 \hat{\bm n}_2 = \eta \hat{\bm n} + s \theta
\hat{\bm \theta}$.
This small-angle approximation is valid because, for one-subhalo term,
we correlate two points in identical subhalos, and their angular
diameter is very small, thus $\theta \ll 1$.
We now have
\begin{eqnarray}
 C_{\rm 1sh}(\theta, \hat{\bm n}) &=& \frac{1}{16\pi^2
 \overline{\VEV{I}}^2} \int ds \int d\eta \int dL
 L^2\frac{dn_{\rm sh}(s\hat{\bm n}, L)}{dL}
 \nonumber\\&&{}\times
 \int \frac{d^3 k}{(2\pi)^3}
 e^{i\bm k\cdot (\eta \hat{\bm n} + s \theta \hat{\bm \theta})}
 |\tilde u(k,M)|^2.
 \label{eq:correlation 1sh 2}
\end{eqnarray}

Angular power spectrum is given by
\begin{equation}
 C_\ell = \frac{1}{2\ell + 1}
  \sum_{m = -\ell}^{\ell} \VEV{|a_{\ell m}|^2}.
  \label{eq:angular power spectrum: definition}
\end{equation}
Using Eq.~(\ref{eq:a_lm squared}) as well as a relation between
spherical harmonics and Legendre polynomials, i.e., Eq.~(46.7) of
Ref.~\cite{Peebles1980}, we have
\begin{eqnarray}
 C_\ell &=& \frac{1}{4\pi f_{\rm sky}^2}
  \int d\Omega_{\hat{\bm n}_1} \int d\Omega_{\hat{\bm n}_2}
  C(\hat{\bm n}_1, \hat{\bm n}_2)
  P_\ell (\hat{\bm n}_1 \cdot \hat{\bm n}_2)
  \nonumber\\
 &\approx& \frac{1}{4\pi f_{\rm sky}^2}
  \int d\Omega_{\hat{\bm n}} \int d \theta 2\pi
  \theta C(\theta, \hat{\bm n}) J_0(\ell \theta)
  \nonumber\\
 &=& \frac{1}{4\pi f_{\rm sky}^2}
  \int d\Omega_{\hat{\bm n}} \int d^2 \theta
  C(\theta, \hat{\bm n}) e^{-i \bm\ell \cdot \bm\theta}.
  \label{eq:power spectrum}
\end{eqnarray}
In the second equality, we used an approximation $P_\ell (\cos\theta)
\approx J_0 (\ell \theta)$ valid for small $\theta$ and large $\ell$'s,
where $P_\ell$ is the Legendre polynomials and $J_0$ is the Bessel
function of zeroth order.
In the last equality, angle integral between $\bm \ell$ and $\bm \theta$
was recovered (see, e.g., Ref.~\cite{Peebles1980}).
This approximation is again particularly good for small angular scales
(large $\ell$'s), where the one-subhalo term would dominate.
Now, it is possible to simplify the $\hat{\bm \theta}$-integral of
$C_{\rm 1sh} e^{-i \bm \ell\cdot \bm \theta}$; with
Eq.~(\ref{eq:correlation 1sh 2}), its relevant part is
\begin{eqnarray}
 \lefteqn{
  \int d^2\theta e^{-i\bm \ell\cdot \bm \theta}
  \int d\eta \int \frac{d^3 k}{(2\pi)^3}
  |\tilde u(k,M)|^2 
  e^{i\bm k\cdot (\eta \hat{\bm n} + s\theta \hat{\bm \theta})}
  }\nonumber\\&=&
  \int d^2 \theta \int d\eta
  \int \frac{dk_\parallel d^2k_\perp}{(2\pi)^3}
  |\tilde u(k,M)|^2 
  e^{ik_\parallel \eta} e^{i\bm \theta \cdot (s \bm k_\perp - \bm \ell)}
  \nonumber\\&=&
  \int dk_\parallel \int d^2 k_\perp |\tilde u(k,M)|^2
  \delta(k_\parallel) \delta^2(s\bm k_\perp - \bm \ell)
  \nonumber\\&=&
  \frac{1}{s^2} \left|\tilde u\left(k = \frac{\ell}{s},
  M\right)\right|^2,
  \label{eq:Limber}
\end{eqnarray}
where in the first equality, we decomposed the wave number $\bm k$ by the
components parallel and perpendicular to $\hat{\bm n}$, i.e., $\bm k =
\bm k_\parallel + \bm k_\perp$, and used $d^3k = dk_\parallel d^2
k_\perp$.

To summarize, combining Eq.~(\ref{eq:power spectrum}) with
Eqs.~(\ref{eq:correlation 1sh 2}) and (\ref{eq:Limber}) for the
one-subhalo term, and with Eq.~(\ref{eq:correlation 2sh}) for the
two-subhalo term, the angular power spectrum of gamma-ray background
from galactic subhalos is
\begin{eqnarray}
 C_\ell &=& C_\ell^{\rm 1sh} + C_\ell^{\rm 2sh},
  \label{eq:angular power spectrum} \\
 C_\ell^{\rm 1sh}
%  &=& \frac{1}{16\pi^2 f_{\rm sky} \overline{\VEV{I}}^2} \int
%  d\Omega_{\hat{\bm n}} \int_{\Lmin}^{\Lmax}dL \int_{s_\ast(L)}^{s_{\rm
%  max}(\hat{\bm n})} \frac{ds}{s^2} L^2
%  \nonumber\\&&{}\times
%  \frac{dn_{\rm sh}(s\hat{\bm n},L)}{dL}
%  \left|\tilde u \left( \frac{\ell}{s}, M\right)\right|^2
%  \nonumber\\
 &\approx& \frac{1}{16\pi^2 f_{\rm sky} \overline{\VEV{I}}^2}
  \int_{\Lmin}^{\Lmax}dL \int_{s_\ast(L)}^{s_{\rm max}}
  \frac{ds}{s^2} L^2
  \nonumber\\&&{}\times
  \frac{d\overline{n_{\rm sh}}(s,L)}{dL}
  \left|\tilde u \left( \frac{\ell}{s}, M\right)\right|^2,
  \label{eq:angular power spectrum 1sh} \\
 C_\ell^{\rm 2sh} &=& \frac{1+\xi_{\rm sh}}{4\pi f_{\rm sky}^2}
  \int d\Omega_{\hat{\bm n}_1} \int d\Omega_{\hat{\bm n}_2}
  \frac{\VEV{I(\hat{\bm n}_1)}\VEV{I(\hat{\bm
  n}_2)}}{\overline{\VEV{I}}^2}
  \nonumber\\&&{}\times
  P_\ell(\hat{\bm n}_1 \cdot \hat{\bm n}_2),
 \label{eq:angular power spectrum 2sh}
\end{eqnarray}
where in Eq.~(\ref{eq:angular power spectrum 1sh}), we again performed
solid-angle integral first, and used angle-averaged luminosity function
$d\overline{n_{\rm sh}}/dL$ in the integrand.
We do not try to further simplify Eq.~(\ref{eq:angular power spectrum
2sh}), as the two-subhalo term would be more important at large angular
scales (as shown below), where the small-angle approximation is no
longer valid.

\section{Results for point-like subhalos}
\label{sec:point source}

In this and subsequent sections, we apply the formulae for angular power
spectrum derived in the previous section to several subhalo models.
Here, first, we consider simple models in which we regard all the
subhalos as gamma-ray point sources, i.e., $u(\bm x, M) = \delta^3 (\bm
x)$.
Its Fourier transform is therefore $\tilde u(k, M) = 1$, independently
of wave number and mass.
Then the one-subhalo term of the angular power spectrum
[Eq.~(\ref{eq:angular power spectrum 1sh})] is independent of multipole
$\ell$, and reduces to the Poisson (shot) noise.

\subsection{Models}
\label{sub:models}

\begin{figure}
\includegraphics[width=8.6cm]{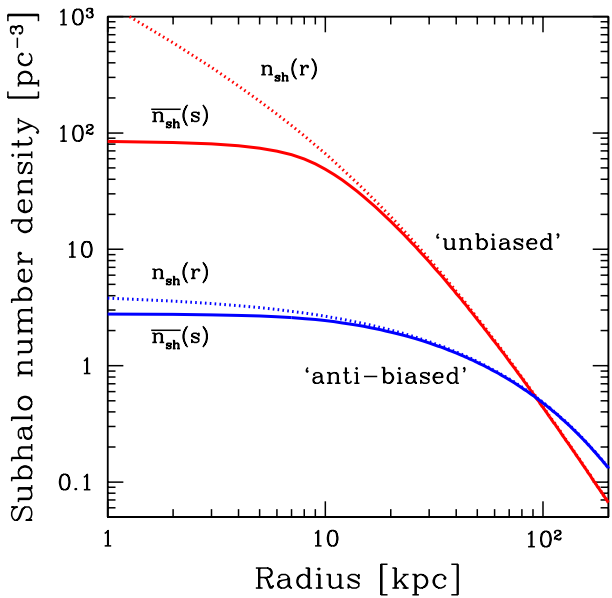}
\caption{Subhalo number densities for unbiased and anti-biased
 distributions, where $\alpha = 2$, $\Mmin = 10^{-6} M_\odot$, $\Mmax =
 10^{10} M_\odot$, and $f = 0.5$.  Densities as a function of
 galactocentric radius, $n_{\rm sh} (r)$, are shown as dotted curves,
 and angle-averaged densities as a function of distance from Earth $s$,
 $\overline{n_{\rm sh}}(s)$ are shown as solid curves.}
\label{fig:density}
\end{figure}

Following results of recent numerical simulations, we adopt the
power-law mass function, $dn_{\rm sh} / dM \propto M^{-\alpha}$, and
assume that mass distribution is independent of subhalo positions in the
host; i.e.,
\begin{equation}
 \frac{dn_{\rm sh}(r, M)}{dM} = n_{\rm sh}(r)
  \frac{\alpha - 1}{\Mmin}
  \left(\frac{M}{\Mmin}\right)^{-\alpha},
  \label{eq:power-law mass function}
\end{equation}
where we assumed $\Mmin \ll \Mmax$ and typical value for $\alpha$ is
about 2.
For the subhalo number density $n_{\rm sh}(r)$, we adopt two different
models.
One is an ``unbiased'' model where the subhalo distribution follows the
NFW density profile of the parent halo, $n_{\rm sh}(r) \propto \rho_{\rm
MW, NFW}(r)$.
The other is an ``anti-biased'' model where the distribution is flatter
than NFW profile and features a central core.
For this model, we adopt the Einasto profile~\cite{Einasto1969},
$n_{\rm sh}(r) \propto \rho_{\rm Ein}(r)$ with parameters $r_{-2} =
0.81 r_{200,{\rm MW}}$ and $\alpha_{\rm E} = 0.68$, where $r_{200,
{\rm MW}}$ is a radius within which the average density is 200 times
the critical density (see Appendix~\ref{app:subhalo density}).
The latter model takes into account the effect of gravitational tidal
disruption of subhalos that is stronger and reduces the number of
subhalos towards the central regions of the halo, and indeed is more
likely to be the case according to numerical
simulations~\cite{Springel2008a} (see also Refs.~\cite{Nagai2005,
Wetzel2008}).
Concrete expressions for $n_{\rm sh}(r)$ for each case are summarized in
Appendix~\ref{app:subhalo density}.
Figure~\ref{fig:density} shows the subhalo number densities $n_{\rm
sh}(r)$, for both the unbiased and anti-biased distributions (dotted
curves).

Figure~\ref{fig:density} also shows the angle-averaged number density
$\overline{n_{\rm sh}}(s)$ (solid).
Note that unlike $n_{\rm sh}(r)$ that is a function of galactocentric
radius $r$, $\overline{n_{\rm sh}}(s)$ is a function of distance from
Earth $s$.
We adopt the same sky region over which we take the average as in
Ref.~\cite{Sreekumar1998}, where the analysis of the isotropic diffuse
gamma-ray background was performed for the Energetic Gamma-Ray
Experimental Telescope (EGRET); i.e., the galactic center
($|b|<30^\circ$ and $|l| < 40^\circ$, where $b$ and $l$ are the galactic
latitude and longitude, respectively) as well as the galactic plane
($|b| < 10^\circ$) are masked, for which $f_{\rm sky} = 0.75$.
As the galactic center is masked, $\overline{n_{\rm sh}}(s)$ is fairly
flat within $\sim$10 kpc, beyond which it reaches regions where the
density drops more rapidly $\rho_{\rm MW} \propto r^{-3}$, and it
eventually becomes almost the same as $n_{\rm sh}(r)$.

\begin{figure}
\includegraphics[width=8.6cm]{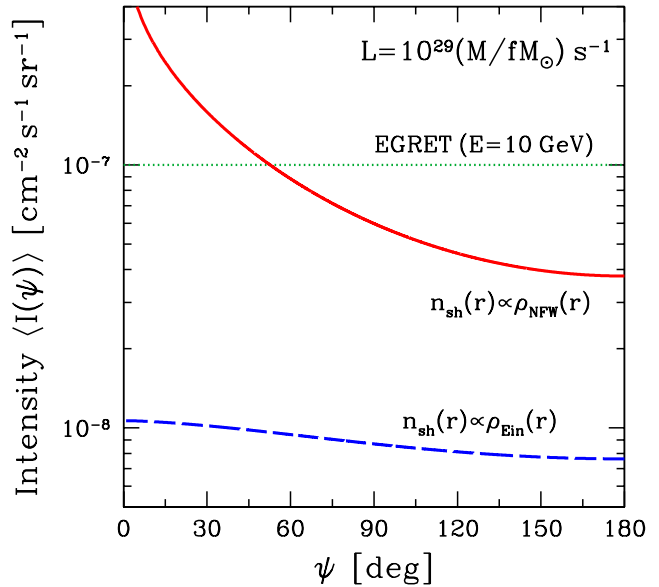}
\caption{Gamma-ray intensity as a function of angle from the galactic
 center $\psi$, for both unbiased (solid) and anti-biased (dashed)
 models.  Subhalo luminosity is related to its mass via $L = 10^{29}
 (M/fM_\odot)$ s$^{-1}$ ($\beta = 1$), and the other parameters are
 $\Mmin = 10^{-6} M_\odot$, $\Mmax = 10^{10} M_\odot$, and $\alpha =
 2$. The isotropic intensity measured with EGRET (for $E = 10$ GeV) is
 shown for comparison (dotted).}
\label{fig:intensity}
\end{figure}

We assume positive correlation between gamma-ray luminosity and mass, $L
\propto M^\beta$ ($\beta > 0$).
Then, the luminosity function is obtained by $dn_{\rm sh}/ dL = (dn_{\rm
sh} / dM)  |dM/dL|$, which yields
\begin{equation}
 \frac{dn_{\rm sh}(r, L)}{dL} = n_{\rm sh}(r)
  \frac{\alpha - 1}{\beta \Lmin}
  \left(\frac{L}{\Lmin}\right)^{-(1-\alpha)/\beta - 1}.
  \label{eq:power-law luminosity function}
\end{equation}
The absolute value of the luminosity, e.g., $\Lmin$, can be kept
arbitrary, because we are here interested in intensity fluctuation {\it
divided} by the mean intensity, where the absolute value cancel
out.\footnote{There is also an implicit dependence on it through the
lower limit of $s$-integral, $s_\ast (L)$, but it is very weak.}
Still we comment that in order to make the mean intensity
$\overline{\VEV{I}}$ as large as the observed value around 10 GeV with
EGRET~\cite{Sreekumar1998}, $\Lmin / \Mmin$ has to be no smaller than
$\sim$10$^{29} f^{-1} M_\odot^{-1}$ s$^{-1}$ if $\beta = 1$ for the
unbiased model~\cite{Ando2008, Lee2008}; for anti-biased model, on the
other hand, this value has to be $\sim$7 times larger.
In Fig.~\ref{fig:intensity}, we show $\VEV{I(\hat{\bm n})}$ as a
function of $\psi$, in the case of $L/M = 10^{29} f^{-1} M_\odot^{-1}$
s$^{-1}$ ($\beta = 1$) for both the unbiased (solid) and anti-biased
(dashed) models.
For comparison, we also show the gamma-ray background intensity measured
with EGRET for 10-GeV photons (dotted)~\cite{Sreekumar1998}.
The number of detectable subhalos and angle-averaged mean intensities
for these models are $\VEV{N_{\rm sh}} = 100$, $\overline{\VEV{I}} = 6.4
\times 10^{-8}$ cm$^{-2}$ s$^{-1}$ sr$^{-1}$ for the unbiased, and
$\VEV{N_{\rm sh}} = 12$, $\overline{\VEV{I}} = 8.7\times 10^{-9}$
cm$^{-2}$ s$^{-1}$ sr$^{-1}$ for the anti-biased distribution.

\subsection{Angular power spectrum}
\label{sub:power spectrum}

We now move on to the angular power spectrum, starting with the
two-subhalo term [Eq.~(\ref{eq:angular power spectrum 2sh})], which
would dominate for large angular scales.
To evaluate $C_\ell^{\rm 2sh}$, we use
HEALPix\footnote{http://healpix.jpl.nasa.gov} package~\cite{Gorski2005}
to generate and analyze the gamma-ray map, for which we used parameters
$N_{\rm side} = 1024$ and $N_{\rm pix} = 12N_{\rm side}^2 \simeq
1.2\times 10^7$ that correspond to pixel size of $0.057^\circ$, small
enough compared with angular resolution of Fermi, $\sim$0.1$^\circ$ for
a 10-GeV photon.
The resulting $C_\ell^{\rm 2sh}$ has a highly oscillatory feature as a
function of $\ell$, so we average over $0.5 \ell$ logarithmic bin for a
given $\ell$.

\begin{figure}
\includegraphics[width=8.6cm]{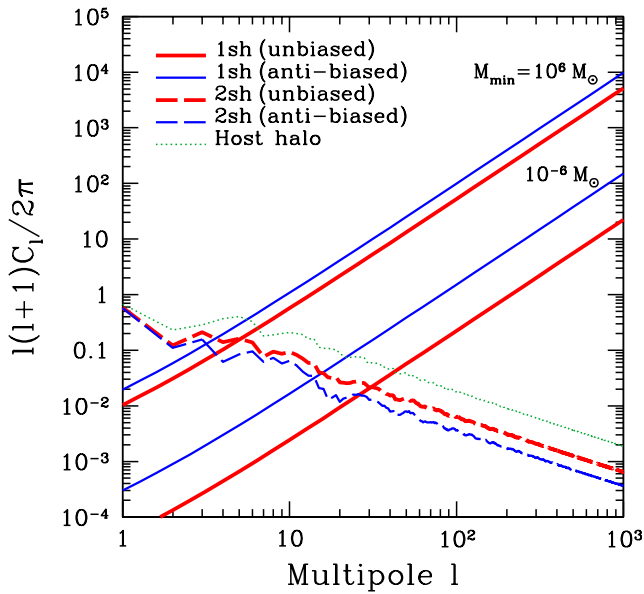}
\caption{Angular power spectrum of the gamma-ray background, where
 subhalos are assumed to be point sources.  Solid curves are for the
 one-subhalo terms $C_\ell^{\rm 1sh}$, dashed for the two-subhalo terms
 $C_\ell^{\rm 2sh}$, and dotted for the host-halo dominated case.  Thick
 (red) and thin (blue) solid/dashed curves are for the unbiased and
 anti-biased subhalo distributions, respectively.  Top and bottom solid
 curves correspond to the cases of $\Mmin = 10^6 M_\odot$ and $10^{-6}
 M_\odot$, respectively, while the other parameters are common $\Mmax =
 10^{10}M_\odot$, $\alpha = 2$, and $\beta = 1$.}
\label{fig:cl_point}
\end{figure}

In Fig.~\ref{fig:cl_point}, we show $\ell(\ell+1) C_\ell^{\rm 2sh} /
2\pi$, for both the unbiased and anti-biased models (dashed curves),
assuming $\xi_{\rm sh} \ll 1$.
As expected, it becomes anisotropic for large angular scales and small
multipole ranges.
At smaller angular scales, where $\ell^2 C_\ell^{\rm 2sh} \propto
\ell^{-1}$, the two-subhalo term becomes less important compared with
the one-subhalo (Poisson) term for which $\ell^2 C_\ell^{\rm 1sh}
\propto \ell^2$ as we see below.
Therefore, in general, the two-subhalo term can be safely neglected for
small angular scales.
The difference between the unbiased and anti-biased distributions is
not very large, a factor of two larger for the former.
We also confirmed that the dependence on a chosen mask is weak.

There is also contribution to the gamma-ray background from a smoothly
distributed dark matter component.
Its emissivity profile is then proportional to $\rho_{\rm MW}^2(r)$ as
annihilation is a two-body process.
The angular power spectrum from this smooth component can be also
evaluated with Eq.~(\ref{eq:angular power spectrum 2sh}), but by using
line-of-sight integral of $\rho_{\rm MW}^2$ for $\VEV{I(\hat{\bm n})}$,
rather than of $\rho_{\rm MW}$ or $\rho_{\rm Ein}$ as we see in, e.g.,
Eq.~(\ref{eq:intensity}).
Figure~\ref{fig:cl_point} also shows the angular power spectrum from the
host-halo component (dotted).
The tendency is the same as the two-subhalo terms, with amplitudes
further larger by a factor of $\sim$3 than the unbiased subhalo
distribution.
Note that in this case, the mean intensity $\overline{\VEV{I}}$ is
evaluated assuming that only the smooth halo gives contribution to the
gamma-ray background (i.e., no substructure).
If we consider both the subhalos and smooth halo, the amplitude is
reduced depending on contribution to the mean intensity from each
component (see Sec.~\ref{sub:results}).

\begin{figure}
\includegraphics[width=8.6cm]{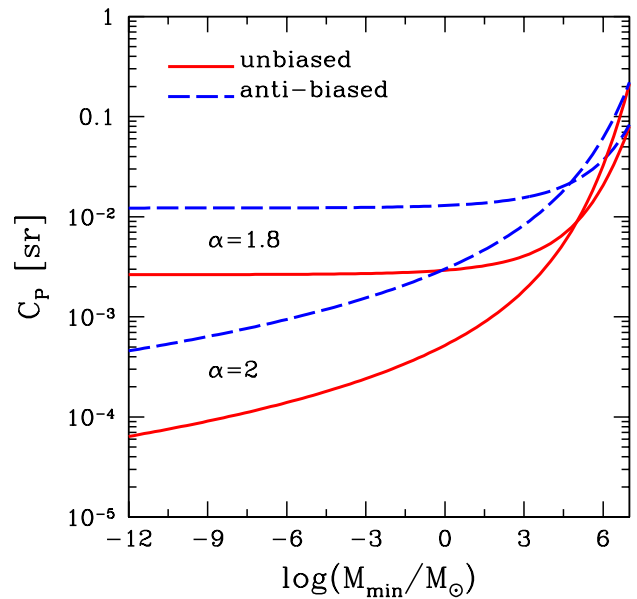}
\caption{The Poisson term of the angular power spectrum $C_P$
 [Eq.~(\ref{eq:Poisson noise})] for unbiased (solid) and anti-biased
 (dashed) distributions.  Top and bottom sets of curves are for $\alpha
 = 1.8$ and 2, respectively.  For the luminosity-mass relation, $L =
 10^{29} (M/fM_\odot)$ s$^{-1}$ is adopted.}
\label{fig:poisson}
\end{figure}

We now discuss the one-subhalo term $C_\ell^{\rm 1sh}$.
If all the subhalos are regarded as point sources, $\tilde u(k,M) = 1$,
$C_\ell^{\rm 1sh}$ becomes independent of $\ell$, $C_\ell^{\rm 1sh} =
C_P$, where the Poisson-noise (shot or white-noise) term is given by
\begin{equation}
 C_P = \frac{1}{16\pi^2 f_{\rm sky} \overline{\VEV{I}}^2}
  \int_{\Lmin}^{\Lmax}dL \int_{s_\ast(L)}^{s_{\rm max}}
  \frac{ds}{s^2} L^2
%  \nonumber\\&&{}\times
  \frac{d\overline{n_{\rm sh}}(s,L)}{dL}.
  \label{eq:Poisson noise}
\end{equation}
Its integrand depends on subhalo number density $\overline{n_{\rm sh}}$,
whereas in the denominator $\overline{\VEV{I}}^2$ appears and each
$\overline{\VEV{I}}$ depends on the subhalo density.
Thus, roughly speaking, $C_P$ is inversely proportional to number of
subhalos that give significant contribution to the mean intensity.
In Fig.~\ref{fig:cl_point}, we also show $\ell (\ell + 1) C_\ell^{\rm
1sh} / 2 \pi$ for the unbiased (thick-red) and anti-biased (thin-blue)
distributions.
By fixing the parameters $\Mmax = 10^{10} M_\odot$, $\alpha = 2$, and
$\beta = 1$, we compare results for $\Mmin = 10^6 M_\odot$ and
$10^{-6} M_\odot$.
In the case of larger $\Mmin$, the mean intensity is dominated by bright
(massive), relatively rare subhalos.
On the other hand, in the case of small $\Mmin$, one includes fainter
subhalos, which increases effective number of sources that contribute to
the mean intensity and thus reduces $C_P$.
This trend is clearly seen in Fig.~\ref{fig:cl_point} and consistent
with the earlier report in Ref.~\cite{Siegal-Gaskins2008}.
In Fig.~\ref{fig:poisson}, we plot $C_P$ as a function of $\Mmin$ for
the unbiased and anti-biased distributions and for $\alpha = 2$ and
1.8.
Here we rescaled $f$ such that contribution to it from the mass range of
$10^6$--$10^{10} M_\odot$ is 0.125; by this, for instance, we get $f =
0.5$ (0.15) for $\alpha = 2$ (1.8) and $\Mmin = 10^{-6} M_\odot$.
When the luminosity function is biased towards high-luminosity range as
in the case of $\alpha = 1.8$, including smaller subhalos has less
significant impact on the mean intensity.
Therefore, the angular power spectrum is fairly flat for $\Mmin < 10
M_\odot$ in this case.

\section{Results for extended subhalo models}
\label{sec:realistic case}

\subsection{Subhalo models}
\label{sub:subhalo models}

We move on to evaluating the angular power spectrum with more realistic
models where angular extension of the gamma-ray intensity profile for
each subhalo is taken into account.
The models we use are the same as those given in Sec.~\ref{sub:models}
except for the source extension $\tilde u(k, M)$ as well as the
luminosity-mass relation that then affects the luminosity function.

As the subhalos are in gravitational potential well of the host,
they are subject to tidal disruption, and therefore, their outer
regions are stripped away.
The inner regions, on the other hand, are more resilient against such
an effect.
Hence it would be a good approximation to assume that the subhalo
density profile is given by a truncated NFW profile:
\begin{equation}
  \rho_{\rm sh}(r_{\rm sh}|M) =
  \left\{\begin{array}{lcc}
  \rho_{\rm NFW}(r_{\rm sh}|M) & \mbox{for} & r_{\rm sh}
  \le r_{\rm cut}, \\
  0 & \mbox{for} & r_{\rm sh} > r_{\rm cut}, \end{array} \right.
  \label{eq:truncated NFW}
\end{equation}
where $r_{\rm cut}$ is a cutoff radius and this is typically much
smaller than the virial radius.
This has been studied extensively in Ref.~\cite{Kazantzidis2004} (see
also Refs.~\cite{Stoehr2002, Hayashi2003, Penarrubia2008}).
%In addition, the effect of tidal stripping of outer regions of
%subhalos can be easily incorporated.
%Assuming instead the Einasto density profile~\cite{Einasto1969} with
%$\alpha_{\rm E}\sim 0.16$, where the outer region of subhalos is
%stripped out and that might give better fit to
%simulations~\cite{Gao2008} (see also Refs.~\cite{,
%}), yields fairly similar conclusions to ours.
%This is because the density profile in the innermost regions, from
%which the majority of the annihilation flux originates, is expected to
%remain intact (see Fig.~A.1 of Ref.~\cite{Siegal-Gaskins2008}).

The gamma-ray luminosity is then given by
\begin{eqnarray}
 L &=& b_{\rm sh}
  \frac{\VEV{\sigma v}}{2}\frac{N_\gamma}{m_\chi^2}
  \int dV_{\rm sh} \rho_{\rm sh}^2(r_{\rm sh}|M)
  \nonumber\\
 &=&  \frac{b_{\rm sh} K}{24\pi}\frac{M^2}{r_s^3} \frac{1-1/(1+c_{\rm
  cut})^3}{[\ln(1+c_{\rm cut})- c_{\rm cut}/(1+c_{\rm cut})]^2},
  \label{eq:subhalo luminosity}
\end{eqnarray}
where $\VEV{\sigma v}$ is thermally averaged annihilation cross section
times relative velocity, $m_\chi$ is WIMP mass, and $N_\gamma$ is number
of gamma-ray photons emitted per annihilation.
In the second equality we define
\begin{eqnarray}
 K &\equiv& \frac{\VEV{\sigma v}N_\gamma}{m_\chi^2}
  \nonumber\\
 &=& K_0 \left(\frac{\VEV{\sigma v}}{3 \times 10^{-26} ~ \mathrm{cm^3 ~
  s^{-1}}}\right) \left(\frac{m_\chi}{100~\mathrm{GeV}}\right)^{-2}
  \left(\frac{N_\gamma}{0.6}\right),
  \nonumber\\
\end{eqnarray}
where we normalize each parameter with typical values often taken in the
literature, and $K_0 = 1.8\times 10^{-30}$ cm$^3$ s$^{-1}$ GeV$^{-2}$.
The value of $\VEV{\sigma v}$ is closely related to the relic density
of dark matter as it determines the abundance of dark matter particles
having survived pair annihilation in the early universe.
For the supersymmetric neutralino, whose mass is around 100~GeV,
$\VEV{\sigma v} = 3\times 10^{-26}$ cm$^3$ s$^{-1}$ is the canonical
value~\cite{Jungman1996}, while a wide range of parameter space is
still allowed~\cite{Bertone2005}.
To obtain $N_\gamma = 0.6$, we used the result of
Ref.~\cite{Bergstrom2001}, where the gamma-ray spectrum as a result of
hadronization and decay of $\pi^0$ has been fitted with a simple
formula.
We also introduce additional boost factor $b_{\rm sh}$ due to internal
structure in the subhalo (see below).

The luminosity also depends on the volume integral of the subhalo
density squared $\rho_{\rm sh}^2(r_{\rm sh})$, which is rewritten in
terms of mass $M$, scale radius $r_s$, and ``concentration''
parameter $c_{\rm cut} \equiv r_{\rm cut} / r_s$ of
subhalos.\footnote{We note that concentration parameter is
conventionally defined as a ratio of virial radius and scale radius.
Here, we use the same terminology also for a different, albeit
similar, quantity $c_{\rm cut}$.}
Both $r_s$ and $c_{\rm cut}$ are functions of subhalo mass.
In order to obtain values of these quantities for a given mass, we
adopt scaling relations among various quantities found in the recent
numerical simulations of Ref.~\cite{Springel2008a}.
More specifically, we adopt the following empirical relations between
$V_{\rm max}$, $r_{\rm max}$ and $M$: $M = 3.37 \times 10^7 M_\odot
(V_{\rm max} / 10~ \mathrm{km ~ s^{-1}})^{3.49}$ and $(V_{\rm max}/H_0
r_{\rm max})^2 = 2.9 \times 10^4 (M/10^8 ~ M_\odot)^{-0.18}$, where
$V_{\rm max}$ is the maximum rotation velocity of the subhalo, $r_{\rm
max}$ is the radius at which the rotation curve hits the maximum, and
$H_0$ is the Hubble constant.
We here postulate that for most of the subhalos investigated in
Ref.~\cite{Springel2008a} the density profile is well approximated by
NFW within $r_{\rm max}$ (i.e., $r_{\rm cut} > r_{\rm max}$) and it is
indeed the case for a sample of subhalos shown in Fig.~22 of
Ref.~\cite{Springel2008a}.
Therefore, we can use the relations between $(V_{\rm max}, r_{\rm
max})$ and $(\rho_s, r_s)$ for the NFW profile, $r_s = r_{\rm max} /
2.163$ and $\rho_s = (4.625/4\pi G) (V_{\rm max} / r_s)^2$, to obtain
$\rho_s$ and $r_s$ as a function of subhalo mass $M$.
Then finally, to get the cutoff radius $r_{\rm cut}$, we require that
the volume integral of Eq.~(\ref{eq:truncated NFW}) equals to $M$.
We find that the concentration parameter $c_{\rm cut}$ is a decreasing
function of $M$, but still larger than 2.163 (i.e., $r_{\rm cut} >
2.163 r_s = r_{\rm max}$) even at the resolution limit of the
simulation, $M = 4 \times 10^4 M_\odot$, confirming that this
procedure gives consistent values for $\rho_s$, $r_s$, and $c_{\rm
cut}$.
These empirical relations do not hold at lower mass regions.
Therefore, we adopt two different approaches, one simply extrapolating
the same relations down to Earth-mass scale $10^{-6} M_\odot$, and the
other assuming there is no contribution from subhalos less massive
than $10^4 M_\odot$.

Recent numerical simulations also tend to imply presence of internal
structure within subhalos, i.e., ``sub-subhalos.''
If we include these sub-subhalos, both the subhalo luminosity and its
spatial profile will change.
For the former, it will give additional boost $b_{\rm sh}$, for which we
adopt $b_{\rm sh} = 5$ (2) for $\alpha = 2$ (1.9 and 1.8) as it is
weakly dependent on subhalo masses~\cite{Kuhlen2008}.
Hence, this effect on the angular power spectrum is expected to be
rather minor.
For the latter, on the other hand, the effect would be more prominent.
If the luminosity is dominated by a sub-subhalo component, it would
change the smooth density-squared intensity profile $u(r_{\rm sh}, M)
\propto \rho_{\rm sh}^2 (r_{\rm sh})$ to the clumpy one that is close to
proportionality with the density, i.e., $u(r_{\rm sh}, M) \propto
\rho_{\rm sh} (r_{\rm sh})$.
Although tidal disruption is likely to convert the sub-subhalo component
into the smooth subhalo component without much changing luminosity from
the latter~\cite{Springel2008b}, we still adopt this model regarding it
as an extreme possibility.
The Fourier transforms of the emissivity profile for both cases are
summarized in Appendix~\ref{app:emissivity}.

Considering all of the recent development summarized above, we here
adopt as our fiducial subhalo model the following characteristics:
(i) power-law mass function $dn_{\rm sh} / dM \propto
M^{-1.9}$~\cite{Springel2008a} with maximum mass of $\Mmax = 10^{10}
M_{\odot}$;
(ii) anti-biased subhalo distribution in the galactic halo, $n_{\rm sh}
\propto \rho_{\rm Ein}$;
(iii) scale radius $r_s$ and concentration parameter $c_{\rm cut}$ for
subhalos as a function of mass obtained from the empirical relations
in Ref.~\cite{Springel2008a};
(iv) subhalo luminosity obtained with canonical values of the
particle-physics parameters, $K = K_0$, and the additional subhalo boost
of $b_{\rm sh} = 2$; and
(v) emissivity profile of each subhalo dominated by its smooth
component, $u \propto \rho_{\rm sh}^2$.
We refer to this fiducial set of parameters as the model ``A.''
However, we still do not know if the phenomenological relations
visited in (iii) above as well as the mass function still hold at
the mass scales smaller than the current resolution limit.
Therefore, we adopt two minimum mass scales: $M_{\rm min} = 10^{-6}
M_\odot$ for the model A1 and $M_{\rm min} = 10^4 M_\odot$ for the
model A2.
In Sec.~\ref{sub:results}, we extensively discuss the angular power
spectrum for these fiducial models, and in Sec.~\ref{sub:other} we
study dependence of results on chosen models and parameter values.

\begin{figure}
\includegraphics[width=8.6cm]{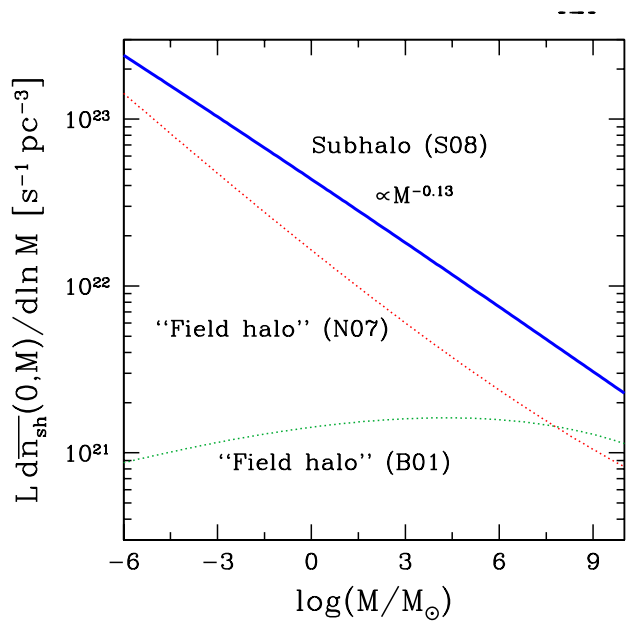}
\caption{Luminosity-weighted mass function for subhalos, $L
  d\overline{n_{\rm sh}}/ d\ln M$ at $s = 0$, following the results of
  the numerical simulation~\cite{Springel2008a} (S08).  The models
  based on calibrations of field halos, Refs.~\cite{Bullock2001} (B01)
  and \cite{Neto2007} (N07) are shown for comparison.}
\label{fig:luminosity}
\end{figure}

\begin{figure}
\includegraphics[width=8.6cm]{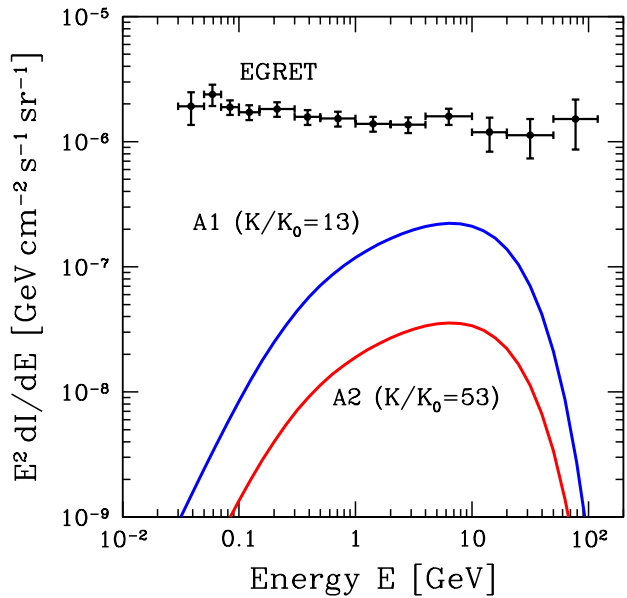}
\caption{Intensity spectrum of the gamma-ray background for subhalo
  models A1 and A2, compared with the EGRET data.  These models are
  boosted by $K/K_0 = 13$ (A1) and 53 (A2).}
\label{fig:spectrum}
\end{figure}

Before closing this subsection, in Fig.~\ref{fig:luminosity}, we show
the luminosity-weighted mass function, $L d\overline{n_{\rm sh}} /
d\ln M$ at $s = 0$ obtained for the fiducial models.
This quantity tells us which mass range contributes to the gamma-ray
intensity the most [see Eq.~(\ref{eq:mean intensity})].
The fiducial subhalo model is labeled as S08 according to the
reference on which this model is based (Ref.~\cite{Springel2008a}),
and this function is well fitted by the $M^{-0.13}$ scaling ($L
\propto M^{0.77}$).
Therefore, the smaller subhalos give more important contribution to
the mean intensity.
In the same figure, we also show results based on other
mass-concentration relations given in Refs.~\cite{Bullock2001} (B01;
see also Refs.~\cite{Kuhlen2005, Maccio2007}) and \cite{Neto2007}
(N07).
These are, however, for field halos that are not fell in
potential well of another larger halo, and hence, their mass is virial
mass and the concentration parameter is defined as the ratio of virial
radius and scale radius as conventionally done.
These models shown in Fig.~\ref{fig:luminosity} thus tell us what the
luminosity-weighted mass function would be {\it if} there were no
tidal forces acting on subhalos.
Note also that these concentration models are calibrated at even
larger scales, such as of galaxies and galaxy clusters, and the
results shown here is based on even more violent extrapolation.
Nevertheless, it is shown that the subhalos are more luminous than the
field halos of equal mass, and this difference might be as large as
two orders of magnitude.
Qualitatively, this difference can be explained as follows.
Tidal force strips the outer region of subhalos away, but the central
region is more strongly bound.
This will reduce the mass of the subhalos significantly but hardly
affect the gamma-ray luminosity that is proportional to the density
squared.
%For example, suppose that a field halo with a virial concentration
%parameter $c_{\rm vir} = 10$ was thrown into another bigger halo
%(became a subhalo), and lost all the masses beyond scale radius
%$r_s$, but was not affected inside at all ($c_{\rm cut} = 1$).
%The mass of the subhalo is now 10\% of the virial mass that this
%subhalo had before merging, but it is still 90\% as bright as it was.
%Therefore, if one compares the luminosity of a subhalo and a field
%halo of {\it same} mass, then the subhalo is more luminous, explaining
%the tendency shown in Fig.~\ref{fig:luminosity}.
Finally, in Fig.~\ref{fig:spectrum}, we show energy spectrum of the
mean intensity $E^2 d\overline{\VEV{I}}/dE$ for the models A1 and A2,
compared with the EGRET data~\cite{Sreekumar1998}.
These subhalo models are boosted by a factor of $K/K_0 = 13$ (A1) and
53 (A2), with which associated anisotropies would be detected (see
discussion in the next subsection).

\subsection{Results for the fiducial model and detectability with Fermi}
\label{sub:results}

\begin{figure}
\includegraphics[width=8.6cm]{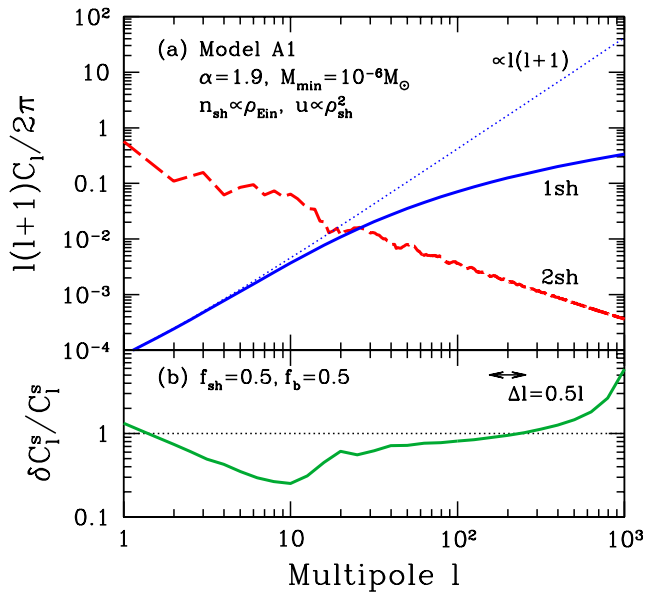}
\caption{(a) Angular power spectrum for the fiducial subhalo model
 with $M_{\rm min} = 10^{-6} M_{\odot}$
 (A1 of Table~\ref{table:models}).  Contributions from the
 one-subhalo and two-subhalo terms are shown as solid and dashed curves,
 respectively, while dotted curve shows Poisson noise that would be
 obtained if subhalos were point sources. (b) Errors for the angular
 power spectrum of the signal $\delta C_\ell^{\rm s} / C_\ell^{\rm s}$,
 for $f_{\rm sh} = 0.5$ and $f_{\rm b} = 0.5$.  The horizontal arrow
 represents the bin width ($\Delta \ell = 0.5 \ell$) for error
 estimates.}
\label{fig:cl_A1}
\end{figure}

In Fig.~\ref{fig:cl_A1}(a), we show $\ell (\ell + 1) C_{\ell} / 2\pi$
for the fiducial model A1.
The two-subhalo term [Eq.~(\ref{eq:angular power spectrum 2sh}) with
$\xi_{\rm sh} \ll 1$] is much smaller than the one-subhalo term
[Eq.~(\ref{eq:angular power spectrum 1sh})] for large multipole
ranges.
For comparison, we also show the Poisson noise [Eq.~(\ref{eq:Poisson
noise})] evaluated for the same model, which would be realized if all
the subhalos were to be gamma-ray point sources.
As expected, the power spectrum is more suppressed at smaller angular
scales (higher multipoles) compared with the noise-like spectrum.
This means that internal structure of the subhalos should be probed with
this analysis.

\begin{figure}
\includegraphics[width=8.6cm]{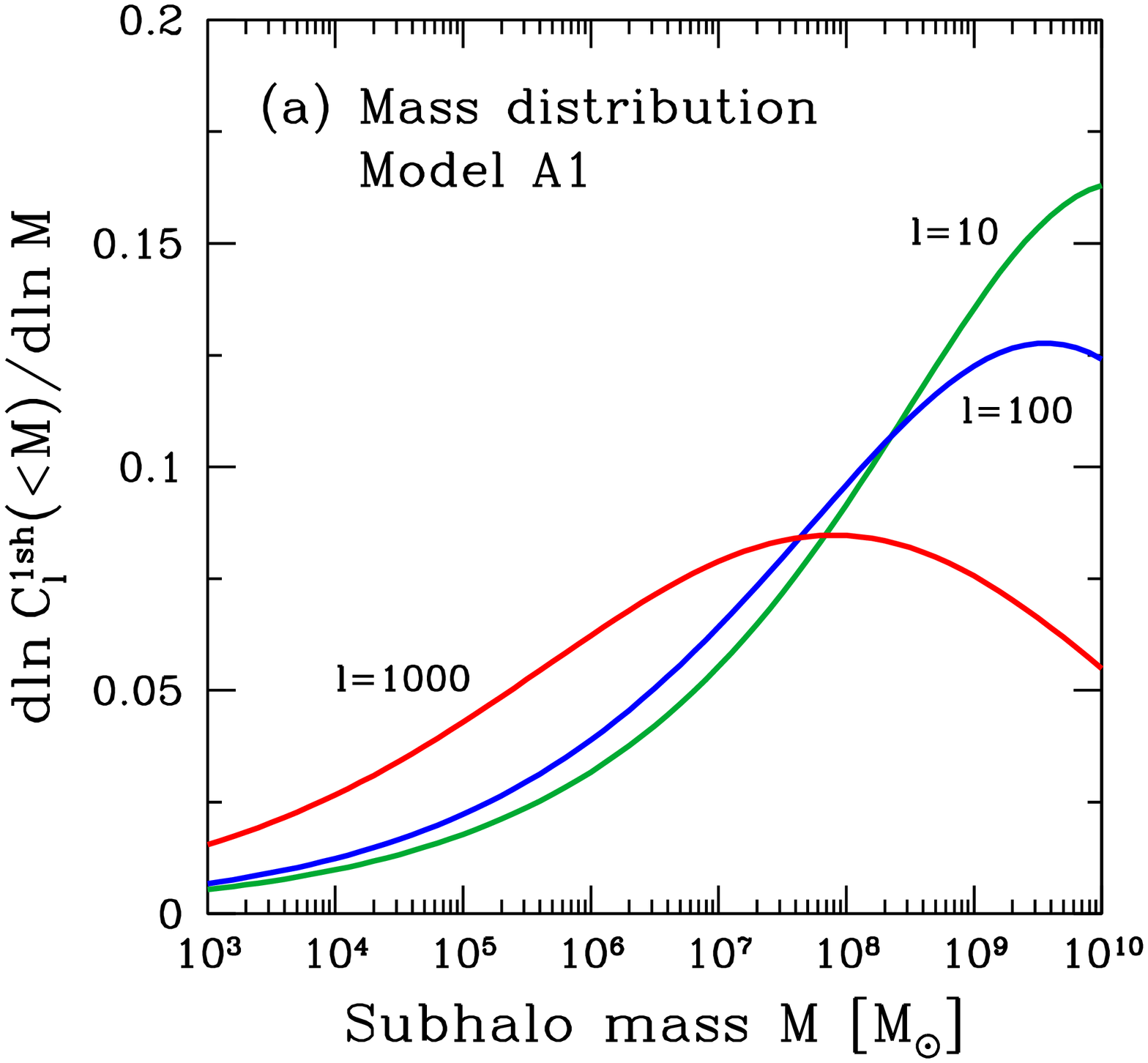}
\includegraphics[width=8.6cm]{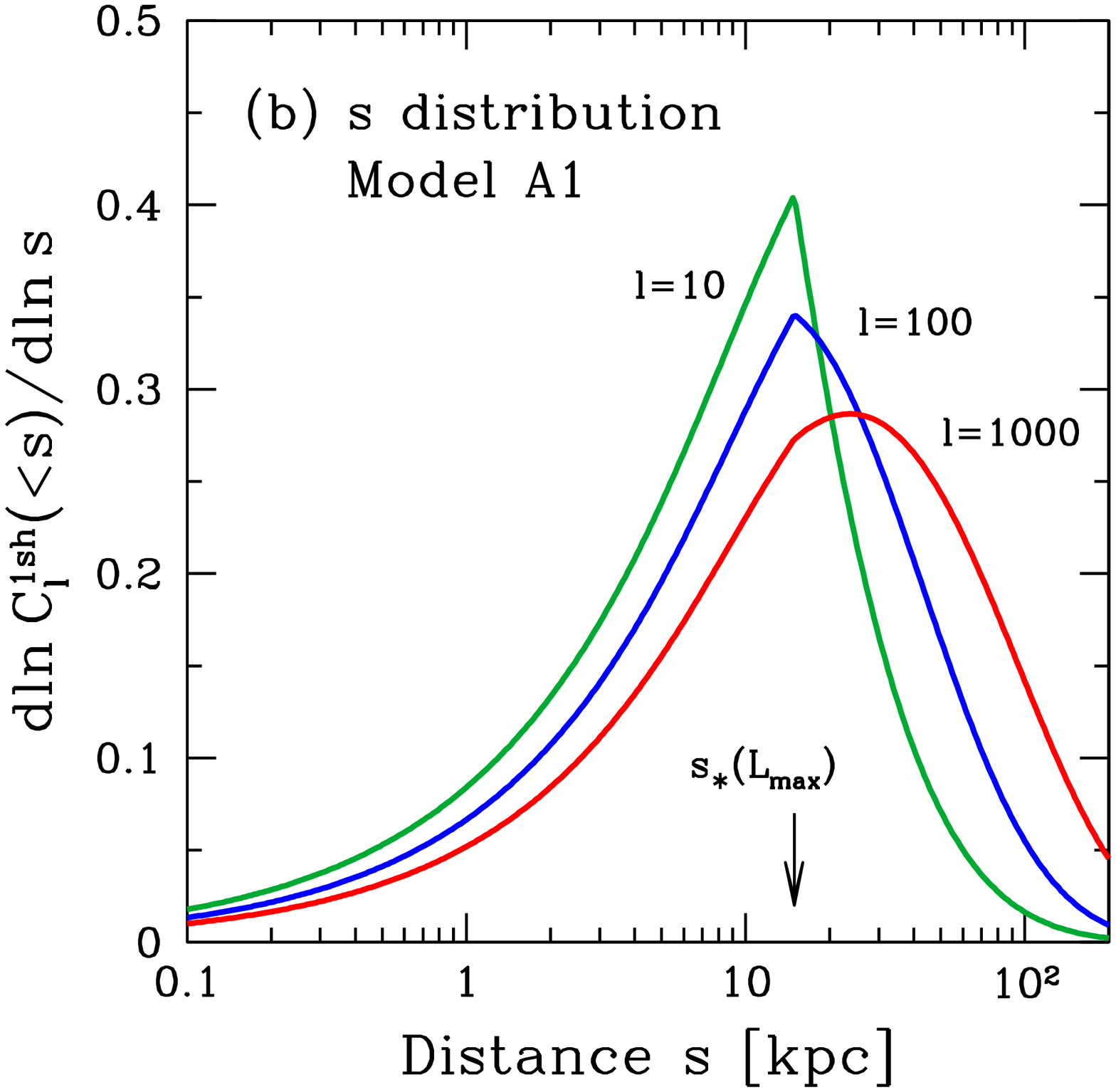}
\caption{Contributions to $C_{\ell}^{\rm 1sh}$ from (a) unit logarithmic
 mass range, and (b) unit logarithmic distance range, for the fiducial
 subhalo model A1.  The curves are for $\ell = 10$, 100, and 1000.}
\label{fig:dist_A1}
\end{figure}

In fact, we can understand this qualitatively, by analyzing the
integrand of Eq.~(\ref{eq:angular power spectrum 1sh}).
In Fig.~\ref{fig:dist_A1}, we show contributions to $C_{\ell}^{\rm
1sh}$ from unit logarithmic mass range and from unit logarithmic
distance ($s$) range.
The mass distributions [Fig.~\ref{fig:dist_A1}(a)] peak at high-mass
range close to $\Mmax$, but are broader for smaller angular scales.
This is because at small angular scales, massive subhalos are regarded
as extended, suppressing the power; note that $|\tilde u(\ell/s,M)|^2$
is a decreasing function of $M$ for fixed $\ell/s$.
Subhalo masses averaged over this distribution and corresponding scale
radii are $1.5\times 10^9 M_{\odot}$ and $r_s = 1.5$ kpc ($\ell =
10$), $1.2 \times 10^9 M_\odot$ and $r_s = 1.4$ kpc ($\ell = 100$), and
$6.4 \times 10^8 M_\odot$ and $r_s = 1.1$ kpc ($\ell = 1000$).
Now, Fig.~\ref{fig:dist_A1}(b) shows that the contribution from farther
subhalos is more important for smaller angular scales, since the closer
subhalos are more extended.
Features at 15 kpc correspond to $s_\ast (\Lmax)$, below which
contribution from massive subhalos are not included as they are
identified as individual sources.
Distances averaged over this distribution are $s = 13$ kpc ($\ell =
10$), 20 kpc ($\ell = 100$), and 32 kpc ($\ell = 1000$).
Combining these typical distance scales with the scale radii, we find
that the angular extension of the subhalos is typically 6.6$^\circ$
($\ell = 10$), 3.9$^\circ$ ($\ell = 100$), and 1.9$^\circ$ ($\ell =
1000$).
For the latter two scales, the subhalo extensions are larger than the
angular scales probed ($\theta \approx 180^\circ / \ell$) and thus
typical subhalos are extended, but for the case of $\ell = 10$, they are
almost point-like sources.
Therefore, as we see in Fig.~\ref{fig:cl_A1}(a), the one-subhalo term
starts to deviate from the white noise above $\ell \sim 10$.

\begin{figure}
\includegraphics[width=8.6cm]{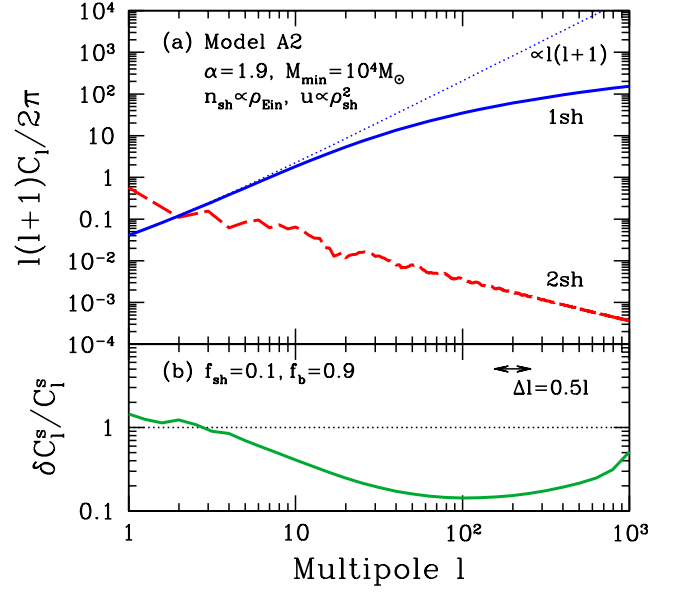}
\caption{The same as Fig.~\ref{fig:cl_A1} but for the fiducial
 model A2 with $M_{\rm min} = 10^4 M_{\odot}$, and (b) $f_{\rm sh}
 = 0.1$ and $f_{\rm b} = 0.9$.}
\label{fig:cl_A2}
\end{figure}

\begin{figure}
\includegraphics[width=8.6cm]{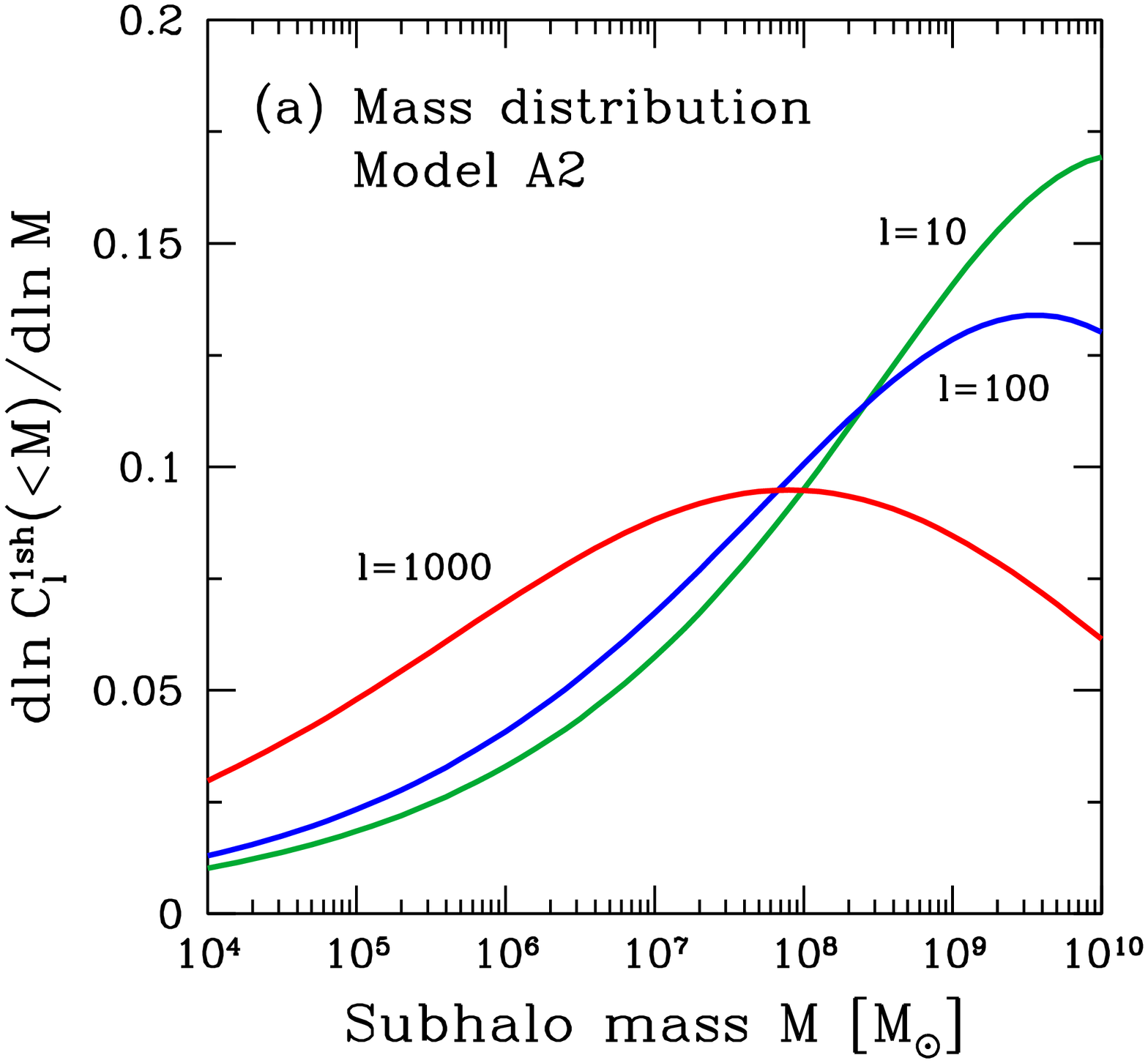}
\includegraphics[width=8.6cm]{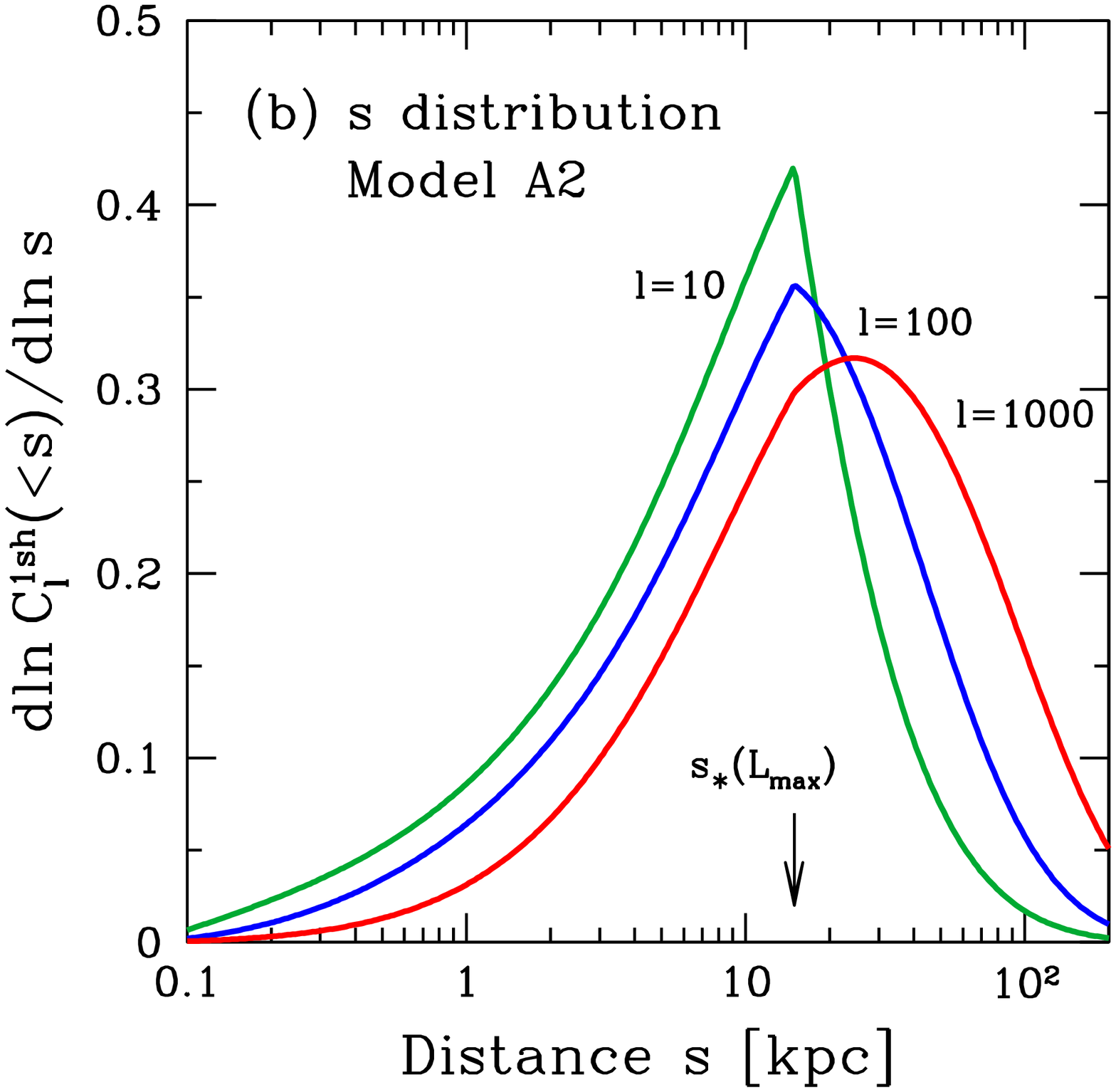}
\caption{The same as Fig.~\ref{fig:dist_A1}, but for the model A2.}
\label{fig:dist_A2}
\end{figure}

In Figs.~\ref{fig:cl_A2}(a) and \ref{fig:dist_A2}, we show the angular
power spectrum, and mass and radius distributions, respectively, for
the other fiducial model A2 ($M_{\rm min} = 10^4 M_\odot$).
The amplitude of the angular power spectrum for the one-subhalo term
is much larger than that for the model A1, whereas the spectrum shape
is almost unchanged.
This dependence and its interpretation are the same as those discussed
in Sec.~\ref{sub:power spectrum} for simplified subhalo models (see
Figs.~\ref{fig:cl_point} and \ref{fig:poisson}).
The mass and distance distributions for $C_\ell^{\rm 1sh}$ are almost
the same as the case of the model A1.

We now discuss the detectability of the angular power spectrum.
The one-sigma errors of $C_\ell$ can be estimated as
\begin{equation}
 \delta C_\ell = \sqrt{\frac{2}{(2\ell+1) \Delta \ell f_{\rm sky}}}
  \left(C_\ell + \frac{C_N}{W_\ell^2}\right),
  \label{eq:error}
\end{equation}
where $\Delta \ell$ is the bin width for that we take $\Delta \ell = 0.5
\ell$, and $W_\ell = \exp(-\sigma_b^2 \ell^2 / 2)$ is the Gaussian
window function with the beam size $\sigma_b = 0.1^\circ$ of the Fermi
(for 10-GeV photons).
The noise power spectrum $C_N$ is associated with the finite photon
count and it is related to number of signal ($N_{\rm s}$) and background
photons ($N_{\rm bg}$) through $C_N = (\Omega_{\rm sky} / N_{\rm s})(1 +
N_{\rm bg} / N_{\rm s})$.
For $N_{\rm s}$ we assume that half of the EGRET gamma-ray background
intensity remains unresolved with Fermi, thus $N_{\rm s} = I_{\rm Fermi}
A_{\rm eff} t_{\rm eff} \Omega_{\rm sky} \approx 1.4 \times 10^5$, where
$I_{\rm Fermi}(E>10~\mathrm{GeV}) = 0.5 I_{\rm EGRET} = 0.5 \times
10^{-7}$ cm$^{-2}$ s$^{-1}$ sr$^{-1}$, $A_{\rm eff} \simeq 10^4$ cm$^2$
is effective area, $t_{\rm eff} = T\Omega_{\rm fov} / 4\pi \approx 3
\times 10^7$ s is effective exposure time, for which we assume 5-yr
all-sky survey ($T = 5$ yr and $\Omega_{\rm fov} = 2.4$ sr is the field
of view of Fermi).
The background due to detector noise is negligible for Fermi, and at
high latitudes, the galactic foreground due to cosmic-ray propagation
will be relatively small compared with the isotropic component of the
gamma-ray background~\cite{Strong2004} (but see also
Ref.~\cite{Keshet2004}).
Hence we approximate $N_{\rm bg} \ll N_{\rm s}$ and obtain $C_N = 6.7
\times 10^{-5}$ sr.

We here consider a multiple-component scenario in which the observed
total gamma-ray intensity comes mainly from several origins.
We assume these are dark matter subhalos, smooth dark matter component
in the host, and an astrophysical source such as blazars:
\begin{equation}
 I_{\rm tot} = I_{\rm sh} + I_{\rm host} + I_{\rm b},
\end{equation}
and we define each fraction by
\begin{equation}
 f_{\rm sh} \equiv \frac{I_{\rm sh}}{I_{\rm tot}}, ~~
  f_{\rm host} \equiv \frac{I_{\rm host}}{I_{\rm tot}},~~
  f_{\rm b} \equiv \frac{I_{\rm b}}{I_{\rm tot}},
\end{equation}
where for simplicity, we represent the angle and ensemble-averaged
intensity $\overline{\VEV{I}}$ by $I$, and the subscript ``b'' stands
for blazars.
Then, the total angular power spectrum corresponding to the total
intensity $I_{\rm tot}$ is given by
\begin{equation}
 C_\ell^{\rm tot} = f_{\rm sh}^2 C_\ell^{\rm sh} + f_{\rm host}^2
 C_\ell^{\rm host} + f_{\rm b}^2 C_\ell^{\rm b} + \mbox{(cross terms)}.
\end{equation}
As galactic dark matter and blazars (or any other extragalactic sources)
are independently distributed in space, cross-correlation terms, $2
f_{\rm sh} f_{\rm b} C_\ell^{\rm sh,b}$ and $2f_{\rm host} f_{\rm b}
C_\ell^{\rm host,b}$ can be safely neglected.
The other cross term $2f_{\rm host}f_{\rm sh} C_\ell^{\rm sh,host}$
will be nonzero, but the amplitude of $C_{\ell}^{\rm sh, host}$ should
be at most comparable to $C_\ell^{\rm host}$ or $C_\ell^{\rm 2sh}$.
Therefore, we shall neglect all the cross-correlation terms in the
following discussions; this is conservative because adding this
additional component would work favorably for dark matter detection.
$C_\ell^{\rm sh}$ is the subhalo angular power spectrum and is the same
as Eq.~(\ref{eq:angular power spectrum}) that we closely investigated
thus far.
The host-halo intensity $I_{\rm host}$ is given by
\begin{equation}
 I_{\rm host} = \frac{(1-f)^2 K}{8\pi}
  \int_0^{r_{\rm vir, MW}} ds \overline{\rho_{\rm MW}^2}(s),
\end{equation}
and for the models A1 and A2, we have $f_{\rm host} / f_{\rm sh} =
6.4\times 10^{-3}$ and 0.16, respectively.
Therefore, the host-halo component would be small in the mean
intensity, and therefore would be further suppressed in the angular
power spectrum.
To estimate the blazar power spectrum $C_\ell^{\rm b}$, we use the
luminosity dependent density evolution model for the gamma-ray
luminosity function~\cite{Narumoto2006, Ando2007b}, and approximate
$C_\ell^{\rm b}$ as nearly a Poisson noise, $C_\ell^{\rm b} \approx
2\times 10^{-3}$ sr, which is good especially for $\ell \agt
30$~\cite{Ando2007a}.
We should use the $C_\ell^{\rm tot}$ in the right-hand side of
Eq.~(\ref{eq:error}) to estimate the errors for subhalo signal
$\delta C_\ell^{\rm sh}$.
Here we are not interested in the blazars and treat them as a reducible
component by using lower-energy data; see Ref.~\cite{Ando2007a} for a
more detailed discussion.

In the following analysis, we take $f_{\rm sh}$ as a free parameter
instead of others such as $b_{\rm sh}$ and $K$.
We neglect the host-halo term as it is always smaller than the subhalo
term, and so $f_{\rm b} = 1 - f_{\rm sh}$.
In Fig.~\ref{fig:cl_A1}(b), we show errors of the angular power
spectrum, $\delta C_\ell^{\rm sh} / f_{\rm sh}^2 C_\ell^{\rm sh}$, for
the model A1, when $f_{\rm sh} = 0.5$ and $f_{\rm b} = 0.5$.
This shows that if subhalos give a fractional contribution to the mean
background intensity, it could also be detected in the angular power
spectrum, especially at $\ell \sim 10$.
Figure~\ref{fig:cl_A2}(b) is the same but for the model A2 and $f_{\rm
sh} = 0.1$ and $f_{\rm b} = 0.9$.
In this case, the detection is more promising as the subhalo
anisotropy is much larger than that of blazars, and signal-to-noise
ratio exceeds 1 for a wide multipole range, reaching maximum at $\ell
\sim 100$.
Therefore, we define detection criterion by setting $\sigma_{\ell} =
1$ at either $\ell = 10$ or 100, where $\sigma_\ell \equiv f_{\rm
sh}^2 C_{\ell}^{\rm sh} / \delta C_{\ell}^{\rm sh}$ is the
signal-to-noise ratio, and for the fiducial models, the value of
$f_{\rm sh}$ required to satisfy this criterion is $f_{\rm sh} =
f_{\rm sh}^{\rm det} = 0.24$
(A1) and 0.038 (A2).
(Although it is for only one-sigma detection, we could also use other
multipole bins to increase significance.)
For the model A1, in order to achieve $f_{\rm sh} = 0.24$, we need
additional boost of $K/K_0 = K_{\rm det} / K_0 = 13$.
For this boost factor, the expected number of subhalo detection is
$\VEV{N_{\rm sh}(K_{\rm det})} = 0.64$.
We also find that associated gamma-ray flux from the galactic center
still satisfies constraint from EGRET~\cite{Mayer-Hasselwander1998},
i.e., $F_{\rm GC}(K_{\rm det}) / F_{\rm GC}^{\rm EGRET} = 0.028$, even
if the $r^{-1}$ cusp of the NFW profile extends to the very center.
Thus, the angular power spectrum could be a stronger probe than
detection as single identified sources.
For the model A2, the values for these quantities associated with the
anisotropy detection are $K_{\rm det} / K_0 = 53$, $\VEV{N_{\rm
sh}(K_{\rm det})} = 4.0$, and $F_{\rm GC}(K_{\rm det}) / F_{\rm
GC}^{\rm EGRET} = 0.12$.
As this model features smaller mean intensity, we need relatively
large boost to give a small fraction ($\sim$4\%) for anisotropy
detection.
Accordingly, the associated number of subhalo detection is a few, but
it is still not very many.
Furthermore, the number of subhalo detection potentially fluctuates to
give $N_{\rm sh} = 0$.
Even in this case, the anisotropy analysis, therefore, might provide
equally sensitive, but statistically more stable method to probe dark
matter annihilation in the galactic substructure.
The spectra of the mean intensity for these models A1 and A2 with the
boost of $K_{\rm det} / K_0$ are shown in Fig.~\ref{fig:spectrum}.

\subsection{Dependence on models and parameters}
\label{sub:other}

\begin{table*}
\begin{center}
\caption{Subhalo models considered in calculations of the angular power
 spectrum.  The first column represents (1) model identification name.
 The fiducial models are A1 and A2.  The second to seventh columns
 specify each model: (2) value of $\Mmin$, (3) $\alpha$, (4) $f$, (5)
 whether subhalo distribution is unbiased or anti-biased ($n_{\rm sh}
 \propto \rho_{\rm NFW}$ or $\rho_{\rm Ein}$), (6) whether
 sub-subhalos dominate luminosity or not ($u \propto \rho_{\rm sh}$ or
 $\rho_{\rm sh}^2$), and (7) the boost factor for subhalos $b_{\rm
 sh}$.   The eighth column shows the values of (8) $f_{\rm host} /
 f_{\rm sh}$. The rest is the values of (9) $f_{\rm sh}^{\rm det}$,
 (10) $K_{\rm det}/K_0$, (11) $\VEV{N_{\rm sh}(K_{\rm det})}$, and
 (12) $F_{\rm GC}(K_{\rm det}) / F_{\rm GC}^{\rm EGRET}$, when the
 subhalo contribution is detected in the angular power spectrum (i.e.,
 $\sigma_{\ell} = f_{\rm sh}^2 C_{\ell}^{\rm sh} / \delta
 C_{\ell}^{\rm sh} = 1$) at either $\ell = 10$ or 100.}
\label{table:models}
\begin{tabular}{lcccccccccccc}\hline\hline
Model & $\Mmin$ & $\alpha$ & $f$ & $n_{\rm sh}$ & $u(r_{\rm sh})$ &
 $b_{\rm sh}$ & & $\frac{f_{\rm host}}{f_{\rm sh}}$ &
 $f_{\rm sh}^{\rm det}$ & $\frac{K_{\rm det}}{K_0}$ & $\VEV{N_{\rm
 sh}(K_{\rm det})}$ & $\frac{F_{\rm GC}(K_{\rm det})}{F_{\rm GC}^{\rm
 EGRET}}$\\
\hline
A1 (fiducial) & $10^{-6}M_\odot$ & 1.9 & 0.2 & $\rho_{\rm Ein}$ & $\rho_{\rm
 sh}^2$ & 2 & & 0.0064 & 0.24 & 13 & 0.64 & 0.028\\
A2 (fiducial) & $10^4 M_\odot$ & 1.9 & 0.16 & $\rho_{\rm Ein}$ & $\rho_{\rm
 sh}^2$ & 2 & & 0.16 & 0.038 & 53 & 4.0 & 0.12\\ \hline
B1 & $10^{-6}M_\odot$ & 1.9 & 0.2 & $\rho_{\rm NFW}$ & $\rho_{\rm
 sh}^2$ & 2 & & $7.9\times 10^{-4}$ & 0.21 & 1.4 & 0.72 & 0.010\\
B2 & $10^4 M_\odot$ & 1.9 & 0.16 & $\rho_{\rm NFW}$ & $\rho_{\rm
 sh}^2$ & 2 & & 0.021 & 0.069 & 13 & 12 & 0.033\\
C1a & $10^{-6}M_\odot$ & 2.0 & 0.5 & $\rho_{\rm Ein}$ & $\rho_{\rm
 sh}^2$ & 5 & & $6.9\times 10^{-5}$ & 0.24 & 0.37 & 0.016 & 0.0014\\
C1b & $10^{-6}M_\odot$ & 1.8 & 0.15 & $\rho_{\rm Ein}$ & $\rho_{\rm
 sh}^2$ & 2 & & 0.068 & 0.090 & 47 & 3.2 & 0.11\\
D2 & $10^4 M_\odot$ & 1.9 & 0.16 & $\rho_{\rm Ein}$ & $\rho_{\rm
 sh}$ & 2 & & 0.16 & 0.11 & 170 & 17 & 0.39\\
\hline\hline
\end{tabular}
\end{center}
\end{table*}

In this subsection, we investigate dependence of results on models and
parameters for subhalos.
In Table~\ref{table:models}, we show models with that we investigate the
dependence on $\alpha$, subhalo distribution (unbiased versus
anti-biased), presence of sub-substructure, as listed in the second to
seventh columns of the table (we fix $\Mmax = 10^{10} M_\odot$).
The first two models are the fiducial models, on which we focused in
the previous subsection.
The next two ``B'' models are for the unbiased subhalo distribution.
Models C1a and C1b is the same as A1 but for different mass function.
Lastly, the model D2 is the same as A2 but the subhalos feature much
more extended emissivity profile, proportional to the density
$\rho_{\rm sh}$, which might be the case if there are a lot of
sub-substructure remaining in the subhalos.

\begin{figure}
\includegraphics[width=8.6cm]{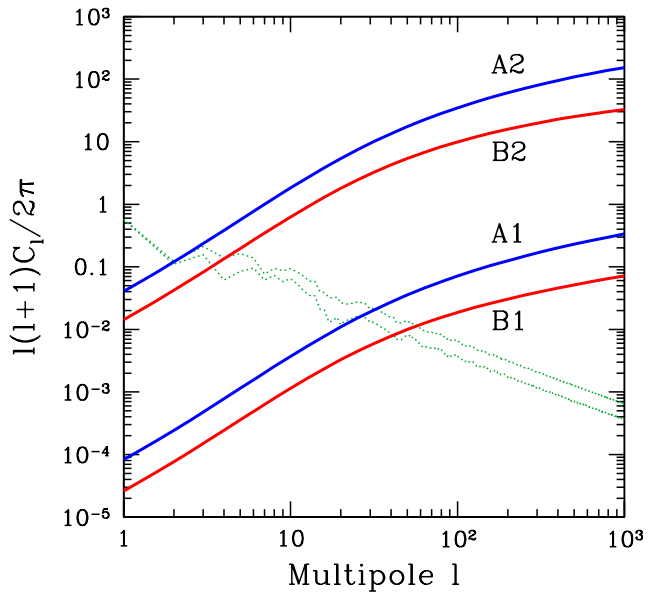}
\caption{Angular power spectrum (one-subhalo term; $C_{\ell}^{\rm 1sh}$)
 of gamma-ray background for the models B1 and B2, compared with A1
 and A2 (Table~\ref{table:models}).  The two-subhalo term is also
 shown as dotted curves for both anti-biased and unbiased subhalo
 distributions.}
\label{fig:cl_B}
\end{figure}

In Fig.~\ref{fig:cl_B}, we show angular power spectrum (one-subhalo
term $C_\ell^{\rm 1sh}$) of B1 and B2 (unbiased), compared with A1 and
A2 (anti-biased).
For all the models investigated, the angular power spectrum is larger
for the anti-biased distribution than the unbiased one.
This tendency is the same as we have seen in the previous section for
point-like subhalos.
Furthermore, deviation from the white noise at small angular scales is
slightly more significant for the unbiased distribution.
This is because the relevant subhalos are closer in the case of the
unbiased model (see Fig.~\ref{fig:density}), and therefore, they are
more extended.
%For example, for the M-ub-sh model, the angular extension (corresponding
%to $r_s$) of typical subhalos is $\sim$9$^\circ$ for $\ell = 100$,
%larger than 4$^\circ$ for the M-ab-sh model.

\begin{figure}
\includegraphics[width=8.6cm]{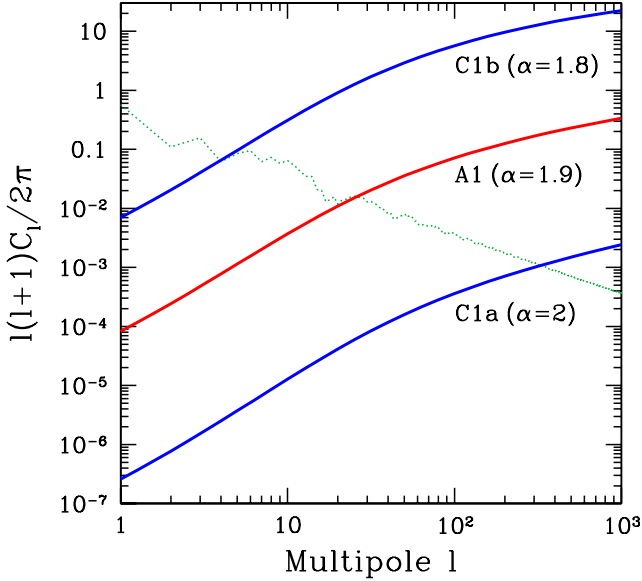}
\caption{Angular power spectrum for the models C1a and C1b, compared
  with A1, to show dependence on $\alpha$.}
\label{fig:cl_C}
\end{figure}

\begin{figure}
\includegraphics[width=8.6cm]{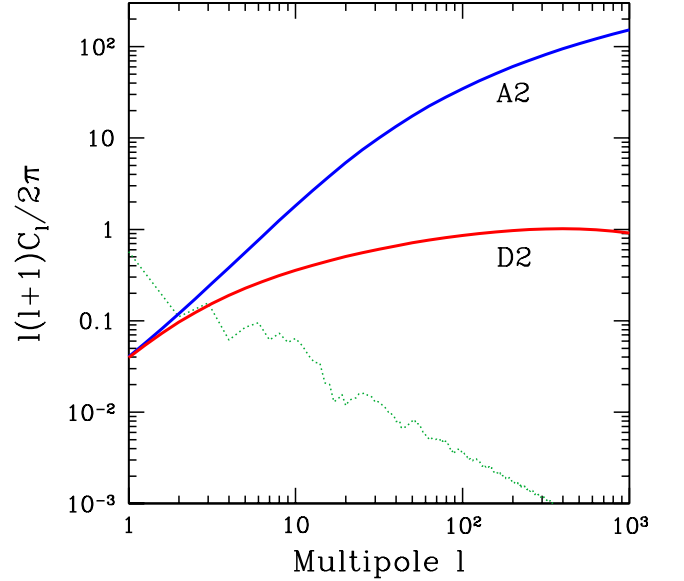}
\caption{Angular power spectrum for the model D2 compared with A2.}
\label{fig:cl_D}
\end{figure}

In Fig.~\ref{fig:cl_C}, we compare the models C1a and C1b with the
fiducial model A1, to study dependence on $\alpha$.
The shape of the power spectrum is almost the same among these models
and the amplitudes follow the same tendency as seen in
Fig.~\ref{fig:poisson}.
Figure~\ref{fig:cl_D} shows a more extended model D2 where we assume
that the gamma-ray luminosity is proportional to the density
$\rho_{\rm sh}$.
The suppression of the power spectrum at small angular scales is even
more prominent for the more extended emissivity profile, which makes
it more difficult for this model to be detected with anisotropy
signals.

Table~\ref{table:models} also shows results of host-to-subhalo ratio for
the mean intensity, $f_{\rm host} / f_{\rm sh}$.
In any of these models, the host-halo component is always smaller than
the subhalo component in the mean intensity.
It is, therefore, even more suppressed in the angular power spectrum.
We also show values of $f_{\rm sh}^{\rm det}$ and the particle-physics
parameter $K_{\rm det}$ (in units of $K_0$) necessary to boost the
subhalo signals to the level of $\sigma_\ell = 1$ where $\ell$ is either
10 or 100.
For any of these models, we find that if the subhalo contribution is
as high as $\sim$20\%, it should be detected in the angular power
spectrum, a similar conclusion as obtained for the extragalactic dark
matter signal~\cite{Ando2007a}.
For the models that feature large $C_\ell^{\rm sh}$ such as A2, B2,
and C1b, the sensitivity of Fermi for $f_{\rm sh}$ could reach down
to several to ten percent level.
We note that to boost the anisotropy signal to the detectable level,
we only need modest values for the particle physics parameter, $K_{\rm
det} / K_0 \sim 1$--100.
The quantities $\VEV{N_{\rm sh}}$ and $F_{\rm GC} / F_{\rm GC}^{\rm
EGRET}$ associated with anisotropy detection are summarized in the
last two columns of Table~\ref{table:models}.
The former ranges from less than one to $\sim$10.
All the models satisfy a constraint from the gamma-ray flux from the
galactic center.

\section{Discussion}
\label{sec:discussion}

Our results confirm what was found in the previous
papers~\cite{Siegal-Gaskins2008, Fornasa2009}; i.e., the angular power
spectrum is dominated by the noise-like term with a stronger suppression
at higher multipoles.
In particular, behaviors of $C_\ell$ for the models where subhalos
luminosity profile is proportional to its density squared, are very
similar to the power spectrum obtained in
Refs.~\cite{Siegal-Gaskins2008, Fornasa2009}.
This is quite natural because these authors also assumed similar
profiles.
Our formulation, not only gives physical explanations of these results,
but also enables to investigate other cases such that the luminosity
profile is simply proportional to density, where the suppression at
small scales is even more pronounced.

For the flux sensitivity of Fermi, we adopted $F_{\rm sens} = 2 \times
10^{-10}$ cm$^{-2}$ s$^{-1}$ for $E > 10$ GeV, but this is only for
sources whose extension is smaller than the beam size (point-like
sources).
If the sources are extended as is the case for massive subhalos, then
the flux sensitivity is worse, and thus $s_\ast (L)$ becomes smaller.
This will decrease expected number of subhalo detection, $\VEV{N_{\rm
sh}}$, because it considerably reduces effective volume.
Thus the conclusions given in the previous section might be rather
conservative, and the anisotropy analysis could be even better,
compared with the detection of subhalos as identified sources.

Throughout our calculations, we took into account the effect of tidal
destruction of subhalos by using the anti-biased distribution, since
stronger gravitational potential around a central region of the host
halo is expected to work more efficiently to get rid of more subhalos.
We also took into account the fact that the subhalos feature higher
concentration than field halos of the same mass.
This might be more complicated because the tidal force is more
effective towards the galactic center, and therefore, the subhalos are
more concentrated there.
This is indeed confirmed by recent numerical simulations, where
concentration parameters scale as a function of galactocentric radius
as $\propto r^{-0.3}$~\cite{Diemand2008}.
Although this effect was not taken into account in our calculations, we
could do that quite easily by using our analytic formulae with slight
modification.

We did not consider dark matter annihilation in the extragalactic halos
in this study, which should also contribute to the isotropic gamma-ray
background.
We note, however, that this component might be comparable to that from
the galactic subhalos~\cite{Oda2005, Fornasa2009}, being smaller by only
a factor of a few.
The anisotropy structure for this component has been investigated in
Refs.~\cite{Ando2006, Ando2007a}, and found that the angular power
spectrum could also be reasonably large, $\ell (\ell + 1) C_\ell^{\rm
ext} / 2 \pi \sim 0.1$ for $\ell \agt 100$~\cite{Ando2007a}.
Including this component further complicates the analysis, but should be
performed when the actual Fermi data became accessible.

Although we focused on gamma-ray photons with $E \agt 10$ GeV, analysis
of angular power spectrum can be performed at any photon energies.
In principle, one could use an energy spectrum of the angular
fluctuation as another diagnosis~\cite{Ando2006, Siegal-Gaskins2009},
although we do not discuss it further.

\section{Conclusions}
\label{sec:conclusions}

To conclude, we investigated the angular power spectrum of the gamma-ray
background from dark matter annihilation in galactic substructure.
Our main findings are the followings.

\begin{enumerate}

\item
In contrast to the earlier works~\cite{Siegal-Gaskins2008, Fornasa2009}
that relied mainly on mock gamma-ray maps generated numerically, we
derived analytic formulae that enable to compute the angular power
spectrum directly with assumed subhalo models
(Sec.~\ref{sec:formulation}).
The angular power spectrum consists of two terms: one-subhalo and
two-subhalo terms [Eq.~(\ref{eq:angular power spectrum})].

\item
The two-subhalo term $C_\ell^{\rm 2sh}$ [Eq.~(\ref{eq:angular power
spectrum 2sh})] depends on smooth (ensemble averaged) distribution
function of subhalos within the parent halo, $n_{\rm sh}$.
This is largely independent of internal density structure of the
subhalos.
We evaluated this term using the HEALPix numerical
package~\cite{Gorski2005} and show that the contribution is considerably
small at small angular scales (e.g., Fig.~\ref{fig:cl_point}).

\item
The one-subhalo term $C_\ell^{\rm 1sh}$ [Eq.~(\ref{eq:angular power
spectrum 1sh})] depends on luminosity profile of each subhalo as well as
number of subhalos that give significant contribution to the mean
intensity of the gamma-ray background.
When the subhalo extension can be neglected (point-like sources), it
simply is the Poisson noise $C_P$ [Eq.~(\ref{eq:Poisson noise})].
In this case, if the subhalo mass function is close to $dn_{\rm sh} / dM
\propto M^{-2}$ as implied by recent numerical simulations, $C_P$
depends on the lower mass cutoff $\Mmin$ as well as whether the subhalos
follow the unbiased or anti-biased distribution in the parent halo
(Fig.~\ref{fig:poisson}).

\item
Taking into account radial extension of the subhalos emissivity profile
suppresses the power spectrum at small angular scales,
through $\tilde u(k, M)$ (Fourier transform of the emissivity profile)
of Eq.~(\ref{eq:angular power spectrum 1sh}).
As fiducial subhalo models, we assume that the emissivity profile
follows internal density squared, as well as the $M^{-1.9}$ mass
spectrum, and the anti-biased subhalo distribution within the host.
We adopt $10^{-6} M_\odot$ and $10^4 M_\odot$ as a minimum mass of the
subhalos.
The angular power spectrum is suppressed compared with the white noise
at angular scales smaller than $\sim$10$^\circ$ because of the
extended intensity profile (Figs.~\ref{fig:cl_A1} and
~\ref{fig:cl_A2}).
If $\Mmin = 10^{-6} M_\odot$, the angular scales around
$\sim$10$^\circ$ is favored for detection, where the power spectrum is
dominated by the two-subhalo term.
For the anisotropy detection, the required fractional contribution to
the mean intensity is $\sim$20\%, for which we need additional
boost of $\sim$10.
The associated number of subhalo detection as individual sources is
less than one.
If $\Mmin = 10^4 M_\odot$, on the other hand, the amplitude of the
angular power spectrum (one-subhalo term) is much larger than the
former case, and therefore, smaller angular scales $\sim$1$^\circ$
would be more promising for detection.
The fractional intensity necessary for anisotropy detection could be
as small as $\sim$4\%, for which the boost is $\sim$50 and the
associated number of subhalo detection is $\sim$4.
Therefore, the analysis of the angular power spectrum might be
stronger than the detection of subhalos as identified gamma-ray
sources, and furthermore, could provide more statistically stable
approach to the same problem.

\item
We also investigated dependence on subhalo parameters by considering
several models (including the fiducial ones) summarized in
Table~\ref{table:models}.
We found that the amplitude of the angular power spectrum is smaller
for the unbiased model compared with the anti-biased model
(Fig.~\ref{fig:cl_B}), and for a softer mass function
(Fig.~\ref{fig:cl_C}).
The angular power spectrum for models featuring more extended
luminosity profile due to dominance by sub-subhalo component is
significantly suppressed at small scales (Figs.~\ref{fig:cl_D}) and is
more difficult to detect with Fermi.

\end{enumerate}
\begin{acknowledgments}
The author is grateful to Eiichiro Komatsu for valuable discussions, and
 thanks Gianfranco Bertone, Samuel Lee, and Jennifer Siegal-Gaskins for
 comments.
This work was supported by the Sherman Fairchild Foundation.
\end{acknowledgments}

\appendix

\section{Number density of subhalos}
\label{app:subhalo density}

Here we obtain number density of subhalos by relating it to the mass of
the Milky-Way halo, for both the unbiased and anti-biased models.
It is obtained by requiring
\begin{equation}
 f M_{\rm vir, MW} = \int_0^{r_{\rm vir, MW}} dr 4\pi r^2
 \int_{\Mmin}^{\Mmax} dM M \frac{dn_{\rm sh}(r,M)}{dM},
\end{equation}
and using Eq.~(\ref{eq:power-law mass function}).
It is also useful to note the following relation:
\begin{equation}
 \frac{M_{\rm vir, MW}}{4\pi r_{s,{\rm MW}}^3 \rho_{s,{\rm MW}}} =
  \ln(1+c_{\rm vir, MW}) - \frac{c_{\rm vir, MW}}{1 + c_{\rm vir, MW}},
\end{equation}
where $c_{\rm vir, MW} \equiv r_{\rm vir, MW} / r_{s, {\rm MW}}$ is the
concentration parameter for the Milky-Way halo.

We thus obtain, for the unbiased case,
\begin{equation}
 n_{\rm sh}(r) =
  \frac{f\rho_{\rm MW}(r)}{\Mmin}\times
  \left\{
   \begin{array}{cc}
    \frac{1}{\ln \Lambda},
     & \mbox{for }\alpha = 2,\\
    \frac{2-\alpha}{\alpha-1} \frac{1}{\Lambda^{2-\alpha}-1},
     & \mbox{for }\alpha \neq 2 .
   \end{array}
  \right.
\end{equation}
For the anti-biased case, we assume that the subhalo distribution
follows Einasto profile~\cite{Einasto1969}, $\rho_{\rm Ein}(r) \propto
\exp [-(2/\alpha_{\rm E}) (r/r_{-2})^{\alpha_{\rm E}}]$, where $r_{-2}$
is a scale radius at which the density slope is $-2$.
With a proper normalization, we obtain
\begin{eqnarray}
 n_{\rm sh}(r) &=&
  \frac{fM_{\rm vir, MW}}{2\pi r_{-2}^3 \Mmin}
%  \nonumber\\&&{}\times
  \gamma\left(\frac{3}{\alpha_{\rm E}}, \frac{2 c_{-2}^{\alpha_{\rm
  E}}}{\alpha_{\rm E}}\right)^{-1}
  \nonumber\\&&{}\times
  \left(\frac{2}{\alpha_{\rm E}}\right)^{3/\alpha_{\rm E}-1}
  \exp\left[- \frac{2}{\alpha_{\rm E}}
 \left(\frac{r}{r_{-2}}\right)^{\alpha_{\rm E}}\right]
 \nonumber\\&&{}\times
  \left\{
   \begin{array}{cc}
    \frac{1}{\ln \Lambda},
     & \mbox{for }\alpha = 2,\\
    \frac{2-\alpha}{\alpha-1} \frac{1}{\Lambda^{2-\alpha}-1},
     & \mbox{for }\alpha \neq 2 ,
   \end{array} \right.
\end{eqnarray}
where $\gamma(a,x)$ is the lower incomplete gamma function, and $c_{-2}
\equiv r_{\rm vir} / r_{-2}$.
For numerical values of the parameters, we adopt $r_{-2} = 0.81
r_{200, {\rm MW}}$ and $\alpha_{\rm E} = 0.68$~\cite{Springel2008a}.

\section{Fourier transform of emissivity profile}
\label{app:emissivity}

The emissivity profile somehow follows density profile of subhalos.
If it is smooth, then $u(r_{\rm sh}, M) \propto \rho_{\rm
sh}^2(r_{\rm sh})$, whereas if they include a number of sub-subhalos,
then it scales more like $\propto \rho_{\rm sh}(r_{\rm sh})$.
Here we compute the Fourier transform of this quantity, $\tilde u(k,
M)$.
When the density profile of the subhalos are well described by the NFW
profile up to a cutoff radius $r_{\rm cut}$, then the both cases have
analytic expressions for $\tilde u(k, M)$ given concentration and
scale radius.
These forms are somewhat complicated, and thus we instead give fitting
formulae that give excellent approximation for a wide range of
reasonable values of $c_{\rm cut}$.
For the both cases, the fitting form is
\begin{equation}
 \tilde u(k,M) = \frac{1}{\left[1+a_1 (kr_s)^{1/b} + a_2
 (kr_s)^{2/b}\right]^b}.
\end{equation}
For $u \propto \rho^2$ (no sub-subhalos), $a_1 = 0.13$, $a_2 = 0$, and
$b = 0.7$, and these values are largely independent of $c_{\rm cut}$.
On the other hand, for $u \propto \rho$ (with sub-subhalos), we have
$a_1 = 100 c_{\rm cut}^{-0.31}$, $a_2 = 170 c_{\rm cut}^{-1.4}$, and $b
= 0.16 c_{\rm cut}^{0.35}$.


\begin{thebibliography}{}


\bibitem{Jungman1996}
  G.~Jungman, M.~Kamionkowski and K.~Griest,
  %``Supersymmetric dark matter,''
  Phys.\ Rep.\  {\bf 267}, 195 (1996)
  [arXiv:hep-ph/9506380].
  %%CITATION = PRPLC,267,195;%%

\bibitem{Hooper2007a}
  D.~Hooper and S.~Profumo,
  %``Dark matter and collider phenomenology of universal extra dimensions,''
  Phys.\ Rep.\  {\bf 453}, 29 (2007)
  [arXiv:hep-ph/0701197].
  %%CITATION = PRPLC,453,29;%%

\bibitem{Bertone2005}
  G.~Bertone, D.~Hooper and J.~Silk,
  %``Particle dark matter: Evidence, candidates and constraints,''
  Phys.\ Rep.\  {\bf 405}, 279 (2005)
  [arXiv:hep-ph/0404175].
  %%CITATION = PRPLC,405,279;%%

\bibitem{Atwood2009}
  F.~W.~B.~Atwood  [LAT Collaboration],
  %``The Large Area Telescope on the Fermi Gamma-ray Space Telescope Mission,''
  Astrophys.\ J.\  {\bf 697}, 1071 (2009)
  [arXiv:0902.1089 [astro-ph.IM]].
  %%CITATION = ASJOA,697,1071;%%

\bibitem{Baltz2008}
  E.~A.~Baltz {\it et al.},
  %``Pre-launch estimates for GLAST sensitivity to Dark Matter annihilation
  %signals,''
  JCAP {\bf 0807}, 013 (2008)
  [arXiv:0806.2911 [astro-ph]].
  %%CITATION = JCAPA,0807,013;%%

\bibitem{Moore1999}
  B.~Moore, S.~Ghigna, F.~Governato, G.~Lake, T.~Quinn, J.~Stadel and P.~Tozzi,
  %``Dark Matter Substructure Within Galactic Halos,''
  Astrophys.\ J.\  {\bf 524}, L19 (1999).
  %%CITATION = ASJOA,524,L19;%%

\bibitem{Klypin1999}
  A.~A.~Klypin, A.~V.~Kravtsov, O.~Valenzuela and F.~Prada,
  %``Where are the missing galactic satellites?,''
  Astrophys.\ J.\  {\bf 522}, 82 (1999)
  [arXiv:astro-ph/9901240].
  %%CITATION = ASJOA,522,82;%%

\bibitem{Diemand2005}
  J.~Diemand, B.~Moore, and J.~Stadel,
  %``Earth-mass dark-matter haloes as the first structures in the early
  %universe,''
  Nature {\bf 433}, 389 (2005)
  [arXiv:astro-ph/0501589].
  %%CITATION = ASTRO-PH 0501589;%%

\bibitem{Diemand2006}
  J.~Diemand, M.~Kuhlen and P.~Madau,
  %``Early supersymmetric cold dark-matter substructure,''
  Astrophys.\ J.\  {\bf 649}, 1 (2006)
  [arXiv:astro-ph/0603250].
  %%CITATION = ASJOA,649,1;%%

\bibitem{Hofmann2001}
  S.~Hofmann, D.~J.~Schwarz and H.~Stoecker,
  %``Damping scales of neutralino cold dark matter,''
  Phys.\ Rev.\  D {\bf 64}, 083507 (2001)
  [arXiv:astro-ph/0104173].
  %%CITATION = PHRVA,D64,083507;%%

\bibitem{Green2005}
  A.~M.~Green, S.~Hofmann and D.~J.~Schwarz,
  %``The first WIMPy halos,''
  JCAP {\bf 0508}, 003 (2005)
  [arXiv:astro-ph/0503387].
  %%CITATION = JCAPA,0508,003;%%

\bibitem{Loeb2005}
  A.~Loeb and M.~Zaldarriaga,
  %``The small-scale power spectrum of cold dark matter,''
  Phys.\ Rev.\  D {\bf 71}, 103520 (2005)
  [arXiv:astro-ph/0504112].
  %%CITATION = PHRVA,D71,103520;%%

\bibitem{Profumo2006}
  S.~Profumo, K.~Sigurdson and M.~Kamionkowski,
  %``What mass are the smallest protohalos?,''
  Phys.\ Rev.\ Lett.\  {\bf 97}, 031301 (2006)
  [arXiv:astro-ph/0603373].
  %%CITATION = PRLTA,97,031301;%%

\bibitem{Bertschinger2006}
  E.~Bertschinger,
  %``The effects of cold dark matter decoupling and pair annihilation on
  %cosmological perturbations,''
  Phys.\ Rev.\  D {\bf 74}, 063509 (2006)
  [arXiv:astro-ph/0607319].
  %%CITATION = PHRVA,D74,063509;%%

\bibitem{Berezinsky1994}
  V.~Berezinsky, A.~Bottino and G.~Mignola,
  %``High-energy gamma radiation from the galactic center due to neutralino
  %annihilation,''
  Phys.\ Lett.\  B {\bf 325}, 136 (1994)
  [arXiv:hep-ph/9402215].
  %%CITATION = PHLTA,B325,136;%%

\bibitem{Bergstrom1998}
  L.~Bergstrom, P.~Ullio and J.~H.~Buckley,
  %``Observability of gamma rays from dark matter neutralino annihilations  in
  %the Milky Way halo,''
  Astropart.\ Phys.\  {\bf 9}, 137 (1998)
  [arXiv:astro-ph/9712318].
  %%CITATION = APHYE,9,137;%%

\bibitem{Cesarini2004}
  A.~Cesarini, F.~Fucito, A.~Lionetto, A.~Morselli and P.~Ullio,
  %``The galactic center as a dark matter gamma-ray source,''
  Astropart.\ Phys.\  {\bf 21}, 267 (2004)
  [arXiv:astro-ph/0305075].
  %%CITATION = APHYE,21,267;%%

\bibitem{Fornengo2004}
  N.~Fornengo, L.~Pieri and S.~Scopel,
  %``Neutralino annihilation into gamma-rays in the Milky Way and in  external
  %galaxies,''
  Phys.\ Rev.\  D {\bf 70}, 103529 (2004)
  [arXiv:hep-ph/0407342].
  %%CITATION = PHRVA,D70,103529;%%

\bibitem{Dodelson2008}
  S.~Dodelson, D.~Hooper and P.~D.~Serpico,
  %``Extracting the Gamma Ray Signal from Dark Matter Annihilation in the
  %Galactic Center Region,''
  Phys.\ Rev.\  D {\bf 77}, 063512 (2008)
  [arXiv:0711.4621 [astro-ph]].
  %%CITATION = PHRVA,D77,063512;%%

\bibitem{Bergstrom1999}
  L.~Bergstrom, J.~Edsjo, P.~Gondolo and P.~Ullio,
  %``Clumpy neutralino dark matter,''
  Phys.\ Rev.\  D {\bf 59}, 043506 (1999)
  [arXiv:astro-ph/9806072].
  %%CITATION = PHRVA,D59,043506;%%

\bibitem{Calcaneo-Roldan2000}
  C.~Calcaneo-Roldan and B.~Moore,
  %``The surface brightness of dark matter: Unique signatures of neutralino
  %annihilation in the galactic halo,''
  Phys.\ Rev.\  D {\bf 62}, 123005 (2000)
  [arXiv:astro-ph/0010056].
  %%CITATION = PHRVA,D62,123005;%%

\bibitem{Tasitsiomi2002}
  A.~Tasitsiomi and A.~V.~Olinto,
  %``The detectability of neutralino clumps via atmospheric Cherenkov
  %telescopes,''
  Phys.\ Rev.\  D {\bf 66}, 083006 (2002)
  [arXiv:astro-ph/0206040].
  %%CITATION = PHRVA,D66,083006;%%

\bibitem{Stoehr2003}
  F.~Stoehr, S.~D.~M.~White, V.~Springel, G.~Tormen and N.~Yoshida,
  %``Dark matter annihilation in the Milky Way's halo,''
  Mon.\ Not.\ R.\ Astron.\ Soc.\  {\bf 345}, 1313 (2003)
  [arXiv:astro-ph/0307026].
  %%CITATION = MNRAA,345,1313;%%

\bibitem{Evans2004}
  N.~W.~Evans, F.~Ferrer and S.~Sarkar,
  %``A 'Baedecker' for the dark matter annihilation signal,''
  Phys.\ Rev.\  D {\bf 69}, 123501 (2004)
  [arXiv:astro-ph/0311145].
  %%CITATION = PHRVA,D69,123501;%%

\bibitem{Aloisio2004}
  R.~Aloisio, P.~Blasi and A.~V.~Olinto,
  %``Gamma-ray constraints on neutralino dark matter clumps in the galactic
  %halo,''
  Astrophys.\ J.\  {\bf 601}, 47 (2004)
  [arXiv:astro-ph/0206036].
  %%CITATION = ASJOA,601,47;%%

\bibitem{Koushiappas2004}
  S.~M.~Koushiappas, A.~R.~Zentner and T.~P.~Walker,
  %``The observability of gamma-rays from neutralino annihilations in Milky  Way
  %substructure,''
  Phys.\ Rev.\  D {\bf 69}, 043501 (2004)
  [arXiv:astro-ph/0309464].
  %%CITATION = PHRVA,D69,043501;%%

\bibitem{Diemand2007a}
  J.~Diemand, M.~Kuhlen and P.~Madau,
  %``Dark matter substructure and gamma-ray annihilation in the Milky Way
  %halo,''
  Astrophys.\ J.\  {\bf 657}, 262 (2007)
  [arXiv:astro-ph/0611370].
  %%CITATION = ASJOA,657,262;%%

\bibitem{Pieri2008}
  L.~Pieri, G.~Bertone and E.~Branchini,
  %``Dark Matter Annihilation in Substructures Revised,''
  Mon.\ Not.\ R.\ Astron.\ Soc.\  {\bf 384}, 1627 (2008)
  [arXiv:0706.2101 [astro-ph]].
  %%CITATION = MNRAA,384,1627;%%

\bibitem{Strigari2008}
  L.~E.~Strigari, S.~M.~Koushiappas, J.~S.~Bullock, M.~Kaplinghat, J.~D.~Simon, M.~Geha and B.~Willman,
  %``The Most Dark Matter Dominated Galaxies: Predicted Gamma-ray Signals from
  %the Faintest Milky Way Dwarfs,''
  Astrophys.\ J.\  {\bf 678}, 614 (2008)
  [arXiv:0709.1510 [astro-ph]].
  %%CITATION = ARXIV:0709.1510;%%

\bibitem{Kuhlen2008}
  M.~Kuhlen, J.~Diemand and P.~Madau,
  %``The Dark Matter Annihilation Signal from Galactic Substructure: Predictions
  %for GLAST,''
  Astrophys.\ J.\  {\bf 686}, 262 (2008)
  [arXiv:0805.4416 [astro-ph]].
  %%CITATION = ARXIV:0805.4416;%%

\bibitem{Martinez2009}
  G.~D.~Martinez, J.~S.~Bullock, M.~Kaplinghat, L.~E.~Strigari and R.~Trotta,
  %``Indirect Dark Matter Detection from Dwarf Satellites: Joint Expectations
  %from Astrophysics and Supersymmetry,''
  arXiv:0902.4715 [astro-ph.HE].
  %%CITATION = ARXIV:0902.4715;%%

\bibitem{Koushiappas2006}
  S.~M.~Koushiappas,
  %``Proper motion of gamma-rays from microhalo sources,''
  Phys.\ Rev.\ Lett.\  {\bf 97}, 191301 (2006)
  [arXiv:astro-ph/0606208].
  %%CITATION = PRLTA,97,191301;%%

\bibitem{Ando2008}
  S.~Ando, M.~Kamionkowski, S.~K.~Lee and S.~M.~Koushiappas,
  %``Can proper motions of dark-matter subhalos be detected?,''
  Phys.\ Rev.\  D {\bf 78}, 101301(R) (2008)
  [arXiv:0809.0886 [astro-ph]].
  %%CITATION = PHRVA,D78,101301;%%

\bibitem{Bergstrom2001}
  L.~Bergstrom, J.~Edsjo and P.~Ullio,
  %``Spectral gamma-ray signatures of cosmological dark matter  annihilations,''
  Phys.\ Rev.\ Lett.\  {\bf 87}, 251301 (2001)
  [arXiv:astro-ph/0105048].
  %%CITATION = PRLTA,87,251301;%%

\bibitem{Ullio2002}
  P.~Ullio, L.~Bergstrom, J.~Edsjo and C.~G.~Lacey,
  %``Cosmological dark matter annihilations into gamma-rays: A closer look,''
  Phys.\ Rev.\  D {\bf 66}, 123502 (2002)
  [arXiv:astro-ph/0207125].
  %%CITATION = PHRVA,D66,123502;%%

\bibitem{Taylor2003}
  J.~E.~Taylor and J.~Silk,
  %``The clumpiness of cold dark matter: Implications for the annihilation
  %signal,''
  Mon.\ Not.\ R.\ Astron.\ Soc.\  {\bf 339}, 505 (2003)
  [arXiv:astro-ph/0207299].
  %%CITATION = MNRAA,339,505;%%

\bibitem{Elsaesser2005}
  D.~Elsaesser and K.~Mannheim,
  %``Supersymmetric dark matter and the extragalactic gamma ray background,''
  Phys.\ Rev.\ Lett.\  {\bf 94}, 171302 (2005)
  [arXiv:astro-ph/0405235].
  %%CITATION = PRLTA,94,171302;%%

\bibitem{Ando2005}
  S.~Ando,
  %``Can dark matter annihilation dominate the extragalactic gamma-ray
  %background?,''
  Phys.\ Rev.\ Lett.\  {\bf 94}, 171303 (2005)
  [arXiv:astro-ph/0503006].
  %%CITATION = PRLTA,94,171303;%%

\bibitem{Oda2005}
  T.~Oda, T.~Totani and M.~Nagashima,
  %``Gamma-ray background from neutralino annihilation in the first
  %cosmological objects,''
  Astrophys.\ J.\  {\bf 633}, L65 (2005)
  [arXiv:astro-ph/0504096].
  %%CITATION = ASJOA,633,L65;%%

\bibitem{Horiuchi2006}
  S.~Horiuchi and S.~Ando,
  %``Dark matter annihilation from intermediate-mass black holes: Contribution
  %to the extragalactic gamma-ray background,''
  Phys.\ Rev.\  D {\bf 74}, 103504 (2006)
  [arXiv:astro-ph/0607042].
  %%CITATION = PHRVA,D74,103504;%%

\bibitem{Ando2006}
  S.~Ando and E.~Komatsu,
  %``Anisotropy of the cosmic gamma-ray background from dark matter
  %annihilation,''
  Phys.\ Rev.\  D {\bf 73}, 023521 (2006)
  [arXiv:astro-ph/0512217].
  %%CITATION = PHRVA,D73,023521;%%

\bibitem{Ando2007a}
  S.~Ando, E.~Komatsu, T.~Narumoto and T.~Totani,
  %``Dark matter annihilation or unresolved astrophysical sources? Anisotropy
  %probe of the origin of cosmic gamma-ray background,''
  Phys.\ Rev.\  D {\bf 75}, 063519 (2007)
  [arXiv:astro-ph/0612467].
  %%CITATION = PHRVA,D75,063519;%%

\bibitem{Cuoco2007}
  A.~Cuoco, S.~Hannestad, T.~Haugbolle, G.~Miele, P.~D.~Serpico and H.~Tu,
  %``The Signature of Large Scale Structures on the Very High Energy   Gamma-Ray
  %Sky,''
  JCAP {\bf 0704}, 013 (2007)
  [arXiv:astro-ph/0612559].
  %%CITATION = JCAPA,0704,013;%%

\bibitem{Cuoco2008}
  A.~Cuoco, J.~Brandbyge, S.~Hannestad, T.~Haugboelle and G.~Miele,
  %``Angular Signatures of Annihilating Dark Matter in the Cosmic Gamma-Ray
  %Background,''
  Phys.\ Rev.\  D {\bf 77}, 123518 (2008)
  [arXiv:0710.4136 [astro-ph]].
  %%CITATION = PHRVA,D77,123518;%%

\bibitem{Zhang2008}
  L.~Zhang and G.~Sigl,
  %``Dark Matter Signatures in the Anisotropic Radio Sky,''
  JCAP {\bf 0809}, 027 (2008)
  [arXiv:0807.3429 [astro-ph]].
  %%CITATION = JCAPA,0809,027;%%

\bibitem{Taoso2008}
  M.~Taoso, S.~Ando, G.~Bertone and S.~Profumo,
  %``Angular correlations in the cosmic gamma-ray background from dark matter
  %annihilation around intermediate-mass black holes,''
  Phys.\ Rev.\  D {\bf 79}, 043521 (2009)
  [arXiv:0811.4493 [astro-ph]].
  %%CITATION = PHRVA,D79,043521;%%

\bibitem{Siegal-Gaskins2008}
  J.~M.~Siegal-Gaskins,
  %``Revealing dark matter substructure with anisotropies in the diffuse
  %gamma-ray background,''
  JCAP {\bf 0810}, 040 (2008)
  [arXiv:0807.1328 [astro-ph]].
  %%CITATION = JCAPA,0810,040;%%

\bibitem{Fornasa2009}
  M.~Fornasa, L.~Pieri, G.~Bertone and E.~Branchini,
  %``Anisotropy probe of galactic and extra-galactic Dark Matter
  %annihilations,''
  arXiv:0901.2921 [astro-ph].
  %%CITATION = ARXIV:0901.2921;%%

\bibitem{Hooper2007b}
  D.~Hooper and P.~D.~Serpico,
  %``Angular Signatures of Dark Matter in the Diffuse Gamma Ray Spectrum,''
  JCAP {\bf 0706}, 013 (2007)
  [arXiv:astro-ph/0702328].
  %%CITATION = JCAPA,0706,013;%%

\bibitem{Yuksel2007}
  H.~Yuksel, S.~Horiuchi, J.~F.~Beacom and S.~Ando,
  %``Neutrino Constraints on the Dark Matter Total Annihilation Cross Section,''
  Phys.\ Rev.\  D {\bf 76}, 123506 (2007)
  [arXiv:0707.0196 [astro-ph]].
  %%CITATION = PHRVA,D76,123506;%%

\bibitem{Lee2008}
  S.~K.~Lee, S.~Ando and M.~Kamionkowski,
  %``The Gamma-Ray-Flux Probability Distribution Function from Galactic Halo
  %Substructure,''
  arXiv:0810.1284 [astro-ph].
  %%CITATION = ARXIV:0810.1284;%%

\bibitem{Dodelson2009}
  S.~Dodelson, A.~V.~Belikov, D.~Hooper and P.~Serpico,
  %``Identifying Dark Matter Annihilation Products In The Diffuse Gamma Ray
  %Background,''
  arXiv:0903.2829 [astro-ph.CO].
  %%CITATION = ARXIV:0903.2829;%%

\bibitem{Navarro1997}
  J.~F.~Navarro, C.~S.~Frenk and S.~D.~M.~White,
  %``A Universal Density Profile from Hierarchical Clustering,''
  Astrophys.\ J.\  {\bf 490}, 493 (1997)
  [arXiv:astro-ph/9611107].
  %%CITATION = ASJOA,490,493;%%

\bibitem{Klypin2002}
  A.~Klypin, H.~Zhao and R.~S.~Somerville,
  %``LCDM-based models for the Milky Way and M31 I: Dynamical Models,''
  Astrophys.\ J.\  {\bf 573}, 597 (2002)
  [arXiv:astro-ph/0110390].
  %%CITATION = ASJOA,573,597;%%

\bibitem{Ghigna2000}
  S.~Ghigna, B.~Moore, F.~Governato, G.~Lake, T.~R.~Quinn and J.~Stadel,
  %``Density profiles and substructure of dark matter halos: converging results
  %at ultra-high numerical resolution,''
  Astrophys.\ J.\  {\bf 544}, 616 (2000)
  [arXiv:astro-ph/9910166].
  %%CITATION = ASJOA,544,616;%%

\bibitem{DeLucia2004}
  G.~De Lucia {\it et al.},
  %``Substructures in Cold Dark Matter Halos,''
  Mon.\ Not.\ Roy.\ Astron.\ Soc.\  {\bf 348}, 333 (2004)
  [arXiv:astro-ph/0306205].
  %%CITATION = MNRAA,348,333;%%

\bibitem{Gao2004}
  L.~Gao, S.~D.~M.~White, A.~Jenkins, F.~Stoehr and V.~Springel,
  %``The subhalo populations of LCDM dark haloes,''
  Mon.\ Not.\ Roy.\ Astron.\ Soc.\  {\bf 355} (2004) 819
  [arXiv:astro-ph/0404589].
  %%CITATION = MNRAA,355,819;%%

\bibitem{Shaw2007}
  L.~Shaw, J.~Weller and J.~P.~Ostriker,
  %``A New Definition of Substructure in Dark Matter Halos,''
  Astrophys.\ J.  {\bf 659}, 1082 (2007)
  [arXiv:astro-ph/0603150].
  %%CITATION = ASTRO-PH/0603150;%%

\bibitem{Diemand2007b}
  J.~Diemand, M.~Kuhlen and P.~Madau,
  %``Formation and evolution of galaxy dark-matter halos and their
  %substructure,''
  Astrophys.\ J.\
  {\bf 667}, 859 (2007)
  [arXiv:astro-ph/0703337].
  %%CITATION = ASJOA,667,859;%%

\bibitem{Seljak2000}
  U.~Seljak,
  %``Analytic model for galaxy and dark matter clustering,''
  Mon.\ Not.\ R.\ Astron.\ Soc.\  {\bf 318}, 203 (2000)
  [arXiv:astro-ph/0001493].
  %%CITATION = MNRAA,318,203;%%

\bibitem{Kravtsov2004}
  A.~V.~Kravtsov, A.~A.~Berlind, R.~H.~Wechsler, A.~A.~Klypin, S.~Gottloeber, B.~Allgood and J.~R.~Primack,
  %``The Dark Side of the Halo Occupation Distribution,''
  Astrophys.\ J.\  {\bf 609}, 35 (2004)
  [arXiv:astro-ph/0308519].
  %%CITATION = ASJOA,609,35;%%

\bibitem{Zentner2005}
  A.~R.~Zentner, A.~A.~Berlind, J.~S.~Bullock, A.~V.~Kravtsov and R.~H.~Wechsler,
  %``The Physics of Galaxy Clustering I: A Model for Subhalo Populations,''
  Astrophys.\ J.\  {\bf 624}, 505 (2005)
  [arXiv:astro-ph/0411586].
  %%CITATION = ASJOA,624,505;%%

\bibitem{Peebles1980}
  P.~J.~E.~Peebles,
  {\it The Large-Scale Structure of the Universe}
  (Princeton University Press, Princeton, NJ, 1980).

\bibitem{Einasto1969}
  J.~Einasto,
  Tr.\ Inst.\ Astrofiz.\ Alma-Ata  {\bf 5}, 87.

\bibitem{Springel2008a}
  V.~Springel {\it et al.},
  %``The Aquarius Project: the subhalos of galactic halos,''
  Mon.\ Not.\ R.\ Astron.\ Soc.\  {\bf 391}, 1685 (2008)
  [arXiv:0809.0898 [astro-ph]].
  %%CITATION = ARXIV:0809.0898;%%

\bibitem{Nagai2005}
  D.~Nagai and A.~V.~Kravtsov,
  %``The Radial Distribution of Galaxies in LCDM clusters,''
  Astrophys.\ J.\  {\bf 618}, 557 (2005)
  [arXiv:astro-ph/0408273].
  %%CITATION = ASJOA,618,557;%%

\bibitem{Wetzel2008}
  A.~R.~Wetzel, J.~D.~Cohn and M.~White,
  %``The Clustering and Host Halos of Galaxy Mergers at High Redshift,''
  arXiv:0810.3650 [astro-ph].
  %%CITATION = ARXIV:0810.3650;%%

\bibitem{Sreekumar1998}
  P.~Sreekumar {\it et al.}  [EGRET Collaboration],
  %``EGRET observations of the extragalactic gamma ray emission,''
  Astrophys.\ J.\  {\bf 494}, 523 (1998)
  [arXiv:astro-ph/9709257].
  %%CITATION = ASJOA,494,523;%%

\bibitem{Gorski2005}
  K.~M.~Gorski, E.~Hivon, A.~J.~Banday, B.~D.~Wandelt, F.~K.~Hansen, M.~Reinecke and M.~Bartelman,
  %``HEALPix -- a Framework for High Resolution Discretization, and Fast
  %Analysis of Data Distributed on the Sphere,''
  Astrophys.\ J.\  {\bf 622}, 759 (2005)
  [arXiv:astro-ph/0409513].
  %%CITATION = ASJOA,622,759;%%

\bibitem{Kazantzidis2004}
  S.~Kazantzidis, L.~Mayer, C.~Mastropietro, J.~Diemand, J.~Stadel and B.~Moore,
  %``Density Profiles of Cold Dark Matter Substructure:Implications for the
  %Missing Satellites Problem,''
  Astrophys.\ J.\  {\bf 608}, 663 (2004)
  [arXiv:astro-ph/0312194].
  %%CITATION = ASJOA,608,663;%%

\bibitem{Stoehr2002}
  F.~Stoehr, S.~D.~M.~White, G.~Tormen and V.~Springel,
  %``The Milky Way's satellite population in a LambdaCDM universe,''
  Mon.\ Not.\ R.\ Astron.\ Soc.\  {\bf 335}, L84 (2002)
  [arXiv:astro-ph/0203342].
  %%CITATION = MNRAA,335,L84;%%

\bibitem{Hayashi2003}
  E.~Hayashi, J.~F.~Navarro, J.~E.~Taylor, J.~Stadel and T.~R.~Quinn,
  %``The Structural Evolution of Substructure,''
  Astrophys.\ J.\  {\bf 584}, 541 (2003)
  [arXiv:astro-ph/0203004].
  %%CITATION = ASJOA,584,541;%%

\bibitem{Penarrubia2008}
  J.~Penarrubia, A.~McConnachie and J.~F.~Navarro,
  %``The cold dark matter halos of local group dwarf spheroidals,''
  Astrophys.\ J.\  {\bf 672}, 904 (2008)  
  [arXiv:astro-ph/0701780].
  %%CITATION = ASTRO-PH/0701780;%%

\bibitem{Springel2008b}
  V.~Springel {\it et al.},
  %``Prospects for detecting supersymmetric dark matter in the Galactic halo,''
  Nature {\bf 456}, 73 (2008)
  [arXiv:0809.0894 [astro-ph]].
  %%CITATION = ARXIV:0809.0894;%%

\bibitem{Bullock2001}
  J.~S.~Bullock {\it et al.},
  %``Profiles of dark haloes: evolution, scatter, and environment,''
  Mon.\ Not.\ R.\ Astron.\ Soc.\  {\bf 321}, 559 (2001)
  [arXiv:astro-ph/9908159].
  %%CITATION = MNRAA,321,559;%%

\bibitem{Kuhlen2005}
  M.~Kuhlen, L.~E.~Strigari, A.~R.~Zentner, J.~S.~Bullock and J.~R.~Primack,
  %``Dark energy and dark matter halos,''
  Mon.\ Not.\ R.\ Astron.\ Soc.\  {\bf 357}, 387 (2005)
  [arXiv:astro-ph/0402210].
  %%CITATION = MNRAA,357,387;%%

\bibitem{Maccio2007}
  A.~V.~Macci{\`o}, A.~A.~Dutton, F.~C.~van den Bosch, B.~Moore,
  D.~Potter and J.~Stadel,
  %``Concentration, Spin and Shape of Dark Matter Haloes: Scatter and the
  %Dependence on Mass and Environment,''
  Mon.\ Not.\ R.\ Astron.\ Soc.\  {\bf 378}, 55 (2007)
  [arXiv:astro-ph/0608157].
  %%CITATION = MNRAA,378,55;%%

\bibitem{Neto2007}
  A.~F.~Neto {\it et al.},
  %``The statistics of LCDM Halo Concentrations,''
  Mon.\ Not.\ R.\ Astron.\ Soc.\  {\bf 381}, 1450 (2007)
  [arXiv:0706.2919 [astro-ph]].
  %%CITATION = ARXIV:0706.2919;%%

\bibitem{Strong2004}
  A.~W.~Strong, I.~V.~Moskalenko and O.~Reimer,
  %``Diffuse Galactic continuum gamma rays. A model compatible with EGRET data
  %and cosmic-ray measurements,''
  Astrophys.\ J.\  {\bf 613}, 962 (2004)
  [arXiv:astro-ph/0406254].
  %%CITATION = ASJOA,613,962;%%

\bibitem{Keshet2004}
  U.~Keshet, E.~Waxman and A.~Loeb,
  %``The Case for a Low Extragalactic Gamma-ray Background,''
  JCAP {\bf 0404}, 006 (2004)
  [arXiv:astro-ph/0306442].
  %%CITATION = JCAPA,0404,006;%%

\bibitem{Narumoto2006}
  T.~Narumoto and T.~Totani,
  %``Gamma-Ray Luminosity Function of Blazars and the Cosmic Gamma-Ray
  %Background: Evidence for the Luminosity Dependent Density Evolution,''
  Astrophys.\ J.\  {\bf 643}, 81 (2006)
  [arXiv:astro-ph/0602178].
  %%CITATION = ASJOA,643,81;%%

\bibitem{Ando2007b}
  S.~Ando, E.~Komatsu, T.~Narumoto and T.~Totani,
  %``Angular power spectrum of gamma-ray sources for GLAST: blazars and clusters
  %of galaxies,''
  Mon.\ Not.\ R.\ Astron.\ Soc.\  {\bf 376}, 1635 (2007)
  [arXiv:astro-ph/0610155].
  %%CITATION = MNRAA,376,1635;%%

\bibitem{Mayer-Hasselwander1998}
  H.~A.~Mayer-Hasselwander {\it et al.},
  %``High-energy gamma ray emission from the galactic center,''
  Astron.\ Astrophys.\  {\bf 335}, 161 (1998).
  %%CITATION = AAEJA,335,161;%%

\bibitem{Diemand2008}
  J.~Diemand, M.~Kuhlen, P.~Madau, M.~Zemp, B.~Moore, D.~Potter and J.~Stadel,
  %``Clumps and streams in the local dark matter distribution,''
  Nature {\bf 454}, 735 (2008)
  [arXiv:0805.1244 [astro-ph]].
  %%CITATION = ARXIV:0805.1244;%%

\bibitem{Siegal-Gaskins2009}
  J.~M.~Siegal-Gaskins and V.~Pavlidou,
  %``Robust identification of isotropic diffuse gamma rays from Galactic dark
  %matter,''
  arXiv:0901.3776 [astro-ph.HE].
  %%CITATION = ARXIV:0901.3776;%%

\end{thebibliography}
\end{document}